\documentclass{article}

\usepackage{arxiv}

\usepackage[english]{babel}

\usepackage[utf8]{inputenc} 
\usepackage[T1]{fontenc}    
\usepackage{amsmath}
\usepackage{graphicx}
  \DeclareGraphicsExtensions{.pdf,.png}
\usepackage{pgfplots}
\pgfplotsset{compat=1.18}
\usepackage{authblk}
\usepackage[colorlinks=true, allcolors=blue]{hyperref}
\usepackage[most]{tcolorbox}
\usepackage{apacite}

\def\BibTeX{{\rm B\kern-.05em{\sc i\kern-.025em b}\kern-.08em
    T\kern-.1667em\lower.7ex\hbox{E}\kern-.125emX}}

\makeatletter
\renewcommand\paragraph{\@startsection{paragraph}{4}{\z@}%
  {3.25ex \@plus1ex \@minus.2ex}%
  {-1em}%
  {\normalfont\normalsize\itshape}}
\makeatother

\begin{document}

\title{Making Sense of the Unsensible: Reflection, Survey, and Challenges for XAI in Large Language Models Toward Human-Centered AI}

\newcommand{\shorttitle}{XAI for LLMs: Survey, Reflection, and Future Challenges} 

\author{ Francisco Herrera$^1$$^2$}

\affil{$^1$Department of Computer Science and Artificial Intelligence, Andalusian Institute of Data Science and Computational Intelligence (DaSCI), University of Granada, Spain. \\ Emails: \texttt{herrera@decsai.ugr.es} }
\affil{$^2$ ADIA Lab, Abu Dhabi,United Arab Emirates}

\date{\today}

\maketitle
\begin{abstract}
As large language models (LLMs) become embedded in sensitive domains such as healthcare, law, and education, the need for transparent, interpretable, and accountable AI systems is more pressing than ever. Explainable AI (XAI) serves as a vital interface between the opaque reasoning of LLMs and the diverse stakeholders who rely on their outputs for high-stakes decisions. This paper offers a comprehensive reflection and survey of XAI for LLMs, organized around three guiding questions: Why is explainability essential? What technical and ethical dimensions does it encompass? And how can it fulfill its role in real-world deployment?

I highlight four core dimensions that shape explainability in LLMs—\textit{faithfulness}, \textit{truthfulness}, \textit{plausibility}, and \textit{contrastivity}—which together expose critical design tensions and inform strategies to generate explanations that are technically robust and contextually meaningful. The paper explores how XAI can support epistemic clarity, regulatory compliance, and user-specific intelligibility across different domains and stakeholder roles.

I also examine how explainability can be measured, in conjunction with key developments in audience-sensitive XAI, mechanistic interpretability, causal reasoning, and adaptive explanation systems. Putting emphasis on the shift from surface-level transparency to governance-ready design, I identify the challenges and research frontiers necessary to ensure the responsible use of LLMs in complex societal contexts. I propose that explainability must evolve into a civic infrastructure that fosters trust, enables contestability, and aligns LLMs with institutional accountability and human-centered decision-making.
\end{abstract}

\begin{keywords} 
~
eXplainable artificial intelligence (XAI), large language models (LLMs), human-centered AI, trustworthy AI, mechanistic interpretability, causal AI, adaptive explanations, audience
\end{keywords}

\newpage

\section*{Highlights}

This paper bridges technical innovation and human-centered governance by surveying and reflecting on XAI strategies for LLMs. It highlights critical challenges and proposes a roadmap towards context-aware, role-calibrated, and actionable explainability infrastructures.

\begin{itemize}
    \item Argues that explainability is a foundational requirement for the ethical and legal deployment of LLMs in high-risk, human-centered domains.
    \item Defines the epistemic and moral roles of XAI, positioning it as both an intelligibility tool and a mechanism of democratic accountability.
    \item Distills four core dimensions of XAI for LLMs: faithfulness, truthfulness, plausibility, and contrastivity, and analyzes their trade-offs and tensions.
    \item Surveys of XAI measures and recent methods, including mechanistic interpretability, causal reasoning, and adaptive explanation systems.
    \item Advocates audience-sensitive and role-calibrated explanation strategies to support trust, contestability, and human-AI collaboration.
    \item Discuss the evolving challenges of XAI in LLMs - from foundational needs and emerging opportunities to the requirements of a governance-ready future - culminating in final reflections on institutional accountability, trust, and responsible deployment.

\end{itemize}

\section*{Graphical abstract}
\begin{figure*}[ht!]
    \centering
    \includegraphics[width=0.70\textwidth]{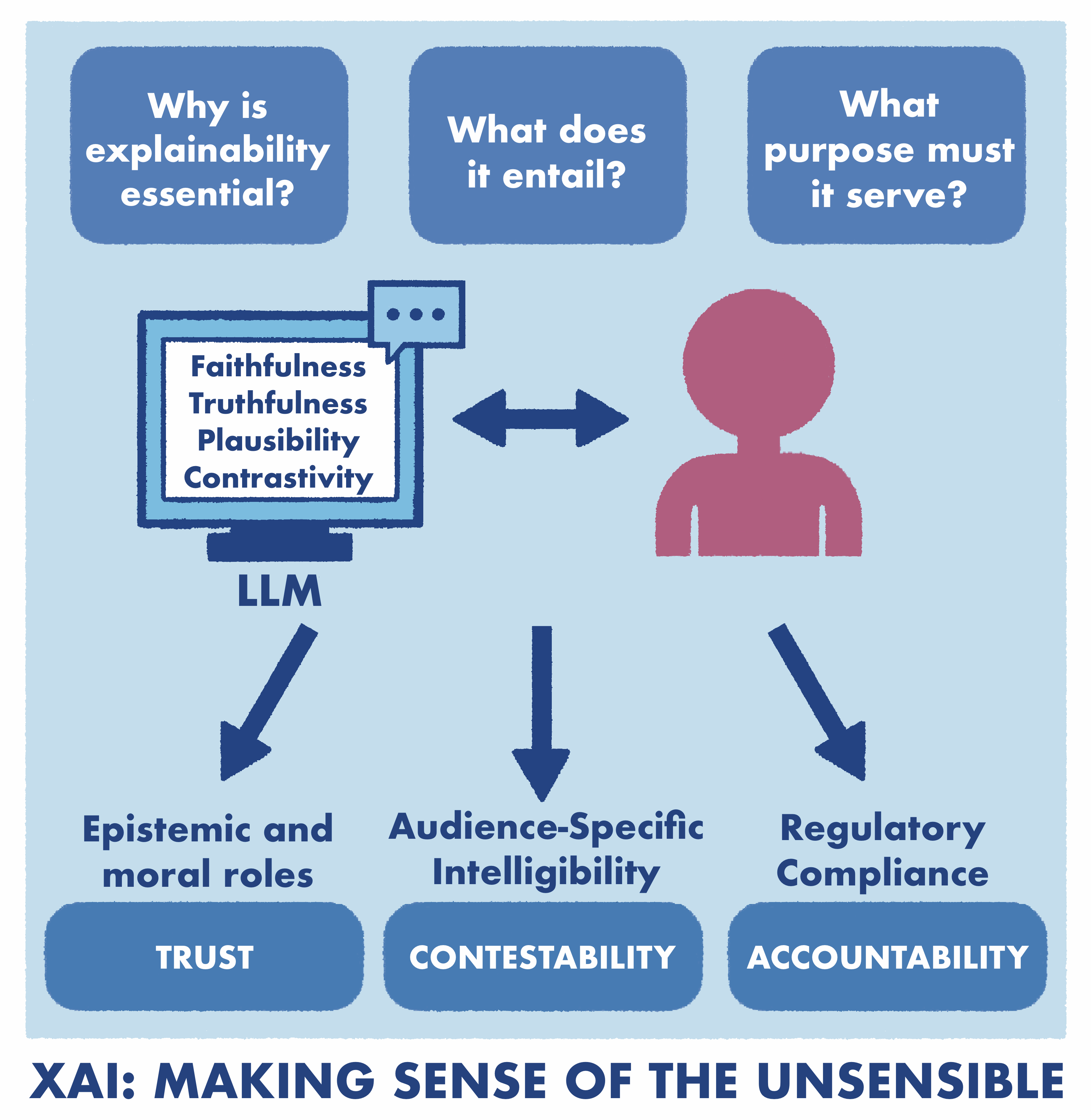}
    \caption{ Reflection, Survey, and Challenges for XAI in Large Language Models Toward Human-Centered AI}
    \label{XAI-LLM}
\end{figure*}

\newpage
\section{Introduction}

The definition and purpose of eXplainable Artificial Intelligence (XAI) are not monolithic; they vary between disciplines, contexts, and stakeholders. As AI systems become more complex, particularly with the advent of large language models (LLM), the need for explainability has evolved significantly. XAI broadly refers to the methods and techniques that make the behavior of AI systems understandable to humans \cite{arrieta2020explainable}:  
\begin{quote}

\noindent {\bf{Definition.}}  \textit{ Given an audience, an explainable AI is one that produces details or reasons to make its functioning clear or easy to understand.}

\end{quote}

Haresamudram et al. 2023 \cite{haresamudram2023three} propose a stakeholder framework that identifies six key audiences for XAI: developers, designers, owners, users, regulators, and society. Each audience has different needs: users want intelligibility and fairness, while regulators require formal justification and auditability.

As AI systems are increasingly being deployed in high-risk sectors such as healthcare, education, law, and finance, among others, the challenges associated with explainability have expanded beyond simple transparency to include ethical, regulatory, and operational concerns among challenges and interdisciplinary research directions \cite{longo2024explainable} . 

XAI is increasingly situated within a broader socio-technical framework. Herrera \cite{herrera2025reflections} emphasizes that explainability is not just about transparency but about enabling actionable understanding. It is a key pillar of trustworthy AI, along with ethics, safety, and accountability. Similarly,  regulatory frameworks such as the EU AI Act\footnote{EU Artificial Intelligence Act, https://artificialintelligenceact.eu/};  the US AI Bill of Rights\footnote{Blueprint for an AI Bill of Rights. 
MAKING AUTOMATED SYSTEMS WORK FOR THE AMERICAN PEOPLE. https://bidenwhitehouse.archives.gov/ostp/ai-bill-of-rights/} and the Artificial Intelligence Law of the People’s Republic of China\footnote{Artificial Intelligence Law of the People’s Republic of China (Draft for Suggestions from Scholars). Translation. https://cset.georgetown.edu/publication/china-ai-law-draft/} demand that AI systems, particularly in sensitive domains, provide intelligible and contestable outputs.  See the blog analysis entitled \textit{"Preparing for compliance: Key differences between EU, Chinese AI regulations"}\footnote{H. Dorwart, H. Qu, T. Bräutigam and J. Gong. Preparing for compliance: Key differences between EU, Chinese AI regulations, Feb. 5, 2025. https://iapp.org/news/a/preparing-for-compliance-key-differences-between-eu-chinese-ai-regulations}

XAI plays a fundamental role in enabling meaningful human-AI collaboration \cite{herrera2025reflections}.  In this view, explanations are designed not only to inform but to empower developers, researchers, and expert regulartors, even expert users, doctors, teachers, to reason with, contest, or validate model outputs within their own domains of expertise. Especially in high-risk, human-centered environments, XAI must facilitate cooperative reasoning and value alignment between AI capabilities and human judgment. It is through this relational lens that XAI becomes essential: not just to decode what models say, but to mediate trust, accountability, and shared decision-making in ethically complex applications in the real world.

The rise of LLMs and other forms of generative or foundation AI (all of them included in the AI Act in the called General Purpose AI (GPAI) models or systems \cite{chairs2025third}, see the paper \cite{triguero2024general} for a broad description) adds a new layer of complexity to explainability. LLMs are particularly prominent and widespread socially and present a new scenario of explainability.

In traditional AI systems, such as those that operate on tabular data or images, the central concern in explainability has been to understand the rationale behind predictions: why a specific outcome was recommended or classified. A variety of explanation techniques (e.g., saliency maps, rule extraction, feature attributions) have been developed to make these systems more transparent. Recently, LLMs themselves have been explored as tools to generate explanations using natural language, offering enhanced accessibility and interpretability LLMs for XAI \cite{hsieh2024comprehensive}. Natural language explanations not only reflect human reasoning, but also make complex AI behavior more comprehensible to diverse stakeholders.

However, when LLMs are the decision-making agents - providing answers, recommendations, or judgments directly - the nature of the explanation becomes more nuanced. These systems already communicate in natural language, yet the mere presence of an answer does not constitute an explanation of its reasoning. The challenge is not just to interpret what was said, but also to understand why it was said. This raises a deeper demand for trustworthy and responsible AI~\cite{diaz2023connecting}. Lin et al. \cite{lin2024towards,wang2024comprehensive} argue that trustworthiness in LLMs requires deliberate interventions to mitigate hallucinations and bias. They advocate for hybrid strategies that incorporate retrieval enhancement, self-refinement, and human-in-the-loop mechanisms to align outputs with societal values and epistemic norms.

In this context, the notion of explanation must be reconsidered. Traditional models often allow for explanation through fixed features or rules. LLMs, in contrast, involve probabilistic text generation, latent representations, and emergent reasoning, factors that complicate how explanation should be structured and evaluated. Explainability in LLMs is therefore multidimensional and addresses technical, epistemic, and communicative goals simultaneously.

To address this complexity, I consider four core dimensions of explainability in LLMs: \textit{faithfulness}, \textit{truthfulness}, \textit{plausibility}, and \textit{contrastivity}. These serve as a conceptual foundation for evaluating and designing explanations that are not only technically aligned with the model's operations but are also comprehensible, contestable, and actionable for users across domains.  These dimensions are examined throughout the work in relation to LLM-specific challenges, tasks, and stakeholder expectations. In the following, I briefly introduce each concept:

\begin{description}

\item[Faithfulness \cite{zhu2024explanation, jacovi2020towards}]  
An explanation is faithful if it accurately reflects the actual reasoning process of the model, that is, the internal mechanisms and representations the model used to arrive at a specific result. This is a model-internal property: it evaluates whether the explanation corresponds to how the model computed the result, not merely whether the explanation is convincing or correct in content. A faithful explanation avoids simplification, fabrication, or misrepresentation of the causal basis of the prediction of the model.

\item[Truthfulness \cite{ji2023survey,lin2021truthfulqa}]
An explanation is truthful if it aligns with external facts or reflects the input of the source in a semantically accurate way. This is an output-level property concerned with factual correctness and the absence of hallucinations, distortions, or omissions. In generative tasks such as summarization or question-answering, a response may be internally consistent, yet untruthful with respect to the original source or known facts. Truthfulness is typically evaluated against ground truth data or curated knowledge bases and is essential for user trust, especially in high-risk domains such as medicine, law, or education.

\item[Plausibility \cite{zhu2024explanation}]  
An explanation is plausible if it is understandable, intuitive, and convincing to its target audience. Plausibility involves naturalness, coherence, and communicative effectiveness from a human perspective. However, LLMs can generate plausible explanations that are not faithful, that is, they appear convincing but do not reflect the true reasoning of the model. This creates a dual tension: plausible explanations may neither reflect the model's internal processes (faithfulness) nor align with external facts or input content (truthfulness).

  \item[Contrastivity \cite{mersha2025unified}] 
  Contrastivity is a critical evaluation that focuses on how well an XAI method can differentiate between different classes through its explanations, providing insight into why a model chooses one class over another. For example, sentiment analysis often distinguishes between positive, neutral, and negative polarities.

\end{description}

The evolution of LLMs has transformed explainability from a technical interpretability challenge into an ethical and governance imperative. This reflection is guided by a central metaphor: \textit{making sense of the unsensible}. In the context of LLMs, this refers to the pressing need to interpret and communicate the behavior of systems whose internal logic is often opaque or emergent. As these models are increasingly deployed in human-centered domains, such as clinical diagnostics, legal analysis, and educational evaluation, the ability to generate explanations that are not only technically grounded but also meaningful to diverse stakeholders becomes essential. In such contexts, XAI is not just a tool for transparency; it constitutes an interpretive infrastructure that supports understanding, contestation, and human-AI collaboration.

I argue that XAI must not only support epistemic goals (clarity, understanding, reasoning), but also uphold moral and institutional functions such as contestability, fairness, and democratic accountability. The rise of LLMs has transformed explainability from a post hoc utility into an essential layer of governance infrastructure. Therefore, drawing on the metaphor of \textit{making sense of the insensible}, this paper proposes that future XAI must bridge the technical opacity of LLM with the lived experiences, rights and roles of various human stakeholders. 

This paper offers a conceptual and normative reflection, rather than a purely empirical or technical study, on the evolving role of explainability in the context of LLMs. Addresses three guiding questions: Why is explainability essential in the LLM era? What does it entail across technical and socioethical dimensions? And what purpose must it serve in real-world deployment? This is complemented with our discussion on XAI measures, the audience-centered role of XAI, and the challenges from the need and opportunities to the real future of governance of XAI in LLMs.

To do that, the paper is organized as follow.  Section~\ref{sec:Why} addresses the question \textit{Why is explainability essential for LLM?}, exploring its epistemic and moral functions, and presenting XAI as a human imperative in light of irreducible and residual opacity. Section~\ref{sec:What} considers the question \textit{What is XAI in the context of LLMs?}, analyzing its role in stakeholder contexts and use cases, and highlighting real-world examples in high-risk domains. Section~\ref{sec:Measure} introduces key evaluation measures for XAI in LLMs, such as faithfulness, truthfulness, plausibility, and contrastivity, and critically examines their limitations and current gaps in benchmarking.  Section Section \ref{sec:Audience} provides a detailed reflection on the role of the audience in shaping explanation strategies and outlines the broader needs and opportunities for audience-sensitive explainability, domain-specific XAI, human-AI collaboration and agentic AI.  In LLMs Section \ref{sec:MI} analyzes a current fundamental aspect for XAI in LLM design, apriori explainability by analyzing the mechanistic interpretability as AI infrastructure, with a tour from modular circuits to cognitive tracing. Section \ref{sec:challenges} discusses challenges and research frontiers in high-risk deployment, from the needs and opportunities of XAI for LLM and towards a governance-ready future for XAI in LLM. 
  Finally, Section~\ref{sec:Conclusion} summarizes the key contributions of the paper and offers concluding reflections on XAI as a foundational infrastructure for socially aligned and governance-ready AI systems.

\section{Beyond Transparency: Why Explainability Is Essential for LLMs?}
\label{sec:Why}

LLMs offer powerful capabilities in language generation and reasoning. However, their architectural opacity, driven by billions of parameters, vast and heterogeneous training data, and emerging behavior, presents significant challenges to accountability, safety, and user trust \cite{zhao2024explainability, luo2024understanding}. As these models are increasingly deployed in high-risk domains, the need for powerful but also interpretable systems has become urgent. In this context, XAI is no longer an optional enhancement or additional - it is a structural necessity to align LLM with human-centered values and institutional accountability.

This section explores explainability beyond surface-level transparency. First, I examine its dual role as an epistemic tool and a moral obligation (Subsection \ref{sec:epistemic-moral}) and how explainability intersects with regulatory frameworks, XAI as a legal and governance obligation (Subsection \ref{sec:Regulatory}). Then I reflect on the imperative for explainability in the face of irreducible and residual opacity (Subsection \ref{sec:opacity}).

\subsection{Epistemic and Moral Foundations of Explainability in LLMs}
\label{sec:epistemic-moral}

The \textit{epistemic function} of explainability lies in its ability to make the inner workings of LLMs intelligible, helping diverse stakeholders - from developers to end-users - understand how and why a model arrives at a particular output. This includes surfacing the data dependencies, decision pathways, or linguistic reasoning involved in a model generation process \cite{arrieta2020explainable, mumuni2025explainable}.

However, explainability is not only about comprehension; it also plays a central \textit{moral} role. As argued in \cite{herrera2025reflections}, it contributes to procedural justice by enabling accountability and ethical governance. In high-risk domains such as healthcare, justice, or education, explanations are not optional; they are essential for challenging and justifying decisions that affect people's lives and rights.

This dual function can be expressed as follows.
\begin{itemize}
    \item \textbf{Epistemically,} explanation enables inquiry: it renders reasoning pathways visible and opens AI behavior to scrutiny and dialogue. In collaborative contexts, for example, medical diagnosis or educational feedback, this supports shared understanding and mutual reasoning between human users and AI agents.
    \item \textbf{Morally,} explanation enables responsibility: equips affected parties with the tools to contest, amend, or affirm decisions. In asymmetric contexts—such as a loan applicant denied credit—transparent explanations are not just informative; they safeguard dignity, autonomy, and the right to challenge outcomes.
\end{itemize}

To meet these functions, XAI must evolve towards \textit{ adaptive, role-sensitive architectures} that dynamically tailor explanations to:
\begin{itemize}
    \item The user’s domain expertise;
    \item The stakes of the decision;
    \item The degree of contestability versus comprehension required.
\end{itemize}

For example, a developer troubleshooting misaligned outputs may need feature attribution and contrastive traces, while a patient using a diagnostic chatbot may benefit from metaphorical or analogy-based explanations. In both cases, the explanation must balance interpretability with practical, ethical, and emotional importance.

This logic aligns with sociotechnical frameworks such as the RED/BLUE XAI model \cite{biecek2024explain}, which differentiates between:
\begin{itemize}
    \item \textbf{RED} \textit{(Research, Explore, Debug)}: Internal, technical explanation for alignment, debugging, and safety auditing;
    \item \textbf{BLUE} \textit{(responsiBle, Legality, trUst, and Ethics)}: External, communicative explanation for users, patients, citizens, and society.
\end{itemize}

Ultimately, XAI is not only a functional tool, but a form of \textit{ethical infrastructure}. It supports transparency, autonomy, and civic trust, values enshrined in regulatory visions like the EU AI Act and the U.S. AI Bill of Rights. As LLMs mediate more decisions, explainability becomes vital not only for understanding but for governance and justice.

\subsection{XAI as a Legal and Governance Obligation}
\label{sec:Regulatory}

Explainability is increasingly recognized not only as a design feature, but as a legal and institutional requirement for responsible deployment. This subsection examines how explainability intersects with regulatory frameworks such as the EU AI Act, GDPR, HIPAA, and the U.S. AI Bill of Rights. It highlights XAI as a legal obligation in high-risk domains and outlines how compliance depends on audience-aware, multilevel intelligibility across technical, institutional, and societal layers.
 
Increasingly, international regulatory frameworks require the transparency, contestability, and intelligibility of AI decisions. The \textit{EU AI Act} classifies many LLM applications as high risk and requires clear documentation of model behavior. The \textit{U.S. AI Bill of Rights} underscores the importance of "notice and explanation" as foundational principles for ethical AI use. It highlights the user’s right to understand AI decisions and challenge their results, particularly in systems that can affect legal status, access to resources, or safety.  Domain-specific laws such as \textit{HIPAA} in healthcare (Health Insurance Portability and Accountability Act)\footnote{Official U.S. government website for HIPAA information: https://www.hhs.gov/hipaa} require systems to protect patient rights, including access to understandable rationales behind clinical decisions. 
 \textit{GDPR} (General Data Protection Regulation)\footnote{GDPR, https://gdpr.eu/} in data protection further reinforce the legal obligation for intelligible algorithmic behavior.  The \textit{GDPR} similarly enshrines a 'right to explanation' for data subjects affected by automated decision making.

As LLMs increasingly permeate high-risk sectors, explainability is no longer a mere technical feature, but a sociolegal imperative. Regulatory bodies across the globe are recognizing the pivotal role of XAI in safeguarding transparency, fairness and accountability AI deployment. Whether in healthcare, finance or public administration, among others,  XAI is being codified as a core requirement to ensure that systems can be audited, contested, and trusted by diverse stakeholders.  Policy developments align with emerging international standards that promote explanation as a prerequisite for human-centered oversight and democratic legitimacy. Regulatory frameworks now emphasize that AI systems must be transparent, fair, and accountable to human users and institutions.

 Effective XAI must be designed with differentiated audiences in mind: technical auditors, regulatory bodies, experts, and end users alike. This multilevel intelligibility is essential for meeting compliance standards and fostering responsible deployment of LLMs. For LLMs operating  explanations are not just technical artifacts—they are governance tools.

\subsection{Managing Residual Opacity: A Human Imperative}
\label{sec:opacity}

Explainability in LLMs is not merely a technical challenge; it is a human imperative. As these systems influence access to services, rights, and opportunities, their outputs must be not only accurate, but comprehensible, and accountable. Whether a patient seeks clarity, a regulator demands justification, or a citizen challenges an automated decision, explainability becomes a foundation for meaningful oversight.

The challenge is compounded by the \textit{opacity} inherent in LLM architectures. These models, trained on massive datasets and composed of billions of parameters, exhibit emergent behavior that resists causal tracing or symbolic analysis. As Amodei \cite{amodei2025interpretability} notes, society faces a "race between interpretability and model intelligence"—a scenario in which our capacity to understand AI systems may lag behind their influence on the world.

Boge \cite{boge2022two} provides a valuable distinction between:
\begin{itemize}
    \item \textbf{Opacity of training:} arising from complex preprocessing, data curation, and hyperparameter tuning—often accidental and reducible.
    \item \textbf{Opacity of representation:} Inherent in how neural networks encode knowledge in weights and activations, irreducible, and intrinsic to the model architecture.
\end{itemize}

This second form, the opacity of representation, is especially challenging in LLM and demands new frameworks for managing, rather than eliminating, opacity.

In this context, explainability must shift from the illusion of full transparency to the goal of ethical scaffolding. This includes:
\begin{itemize}
    \item Designing role-sensitive explanations that prioritize what matters, for whom it matters;
    \item Embedding governance tools like audit logs, decision pathways, and explanation metadata;
    \item Ensuring institutional recourse mechanisms are tied to intelligible system behavior.
\end{itemize}

For example, even when a model generates fluent output, these may include hallucinations or subtle biases that can mislead users in critical contexts, such as clinical triage or legal adjudication. In these scenarios, XAI is not meant to increase raw performance, but to foster trust calibration, enable correction, and preserve procedural justice.

As Fraser et al.~\cite{fraser2022ai} argue, opacity in AI systems is not inherently unethical; what matters is whether it is governed. Rather than treating opacity in LLMs as a flaw to be eliminated, it should be recognized as a structural condition that demands ethical and institutional oversight. Explainability, in this context, is not a panacea nor a promise of full transparency, but a mediating infrastructure that supports contestability, trust calibration, and procedural accountability based on the credibility of the institutions that govern these systems. Instead of striving for total transparency, which may be technically infeasible or epistemically misleading, we advocate for layered, role-sensitive explanations that align with human values, legal mandates, and domain-specific risks.

Rather than treating opacity as a failure, we must manage it responsibly. This means accepting that explainability is not about making everything visible - but about making decisions \textit{contestable, auditable, and fair}. The irreducible opacity of LLMs can be ethically manipulated through transparency strategies, stakeholder-centered design, and governance ethics. In the paper \cite{Herrrera-Reyes2025}, under title \textit{"Opacity as a Feature, Not a Flaw: The LoBOX Governance Ethic for Role-Sensitive Explainability and Institutional Trust in AI"}, the authors argue that opacity is not a flaw to eliminate, but a structural condition to ethically govern. The critical question is not whether we can fully see inside the machine, but whether affected stakeholders can challenge outcomes and trust the institutions that oversee them. This paper has argued that opacity in AI systems, particularly LLMs, should not be seen merely as a technical defect but as a sociotechnical reality that demands responsible design. They propose a framework to manage opacity and explanations for a governable ethic framework. 

In doing so, we offer practical scaffolds for AI developers, ethicists, and policymakers who work to govern opacity through accountable design. These scaffolds include audience-calibrated explanation architectures, multidimensional evaluation grounded in faithfulness, truthfulness, plausibility, and contrastivity, and alignment with governance mechanisms such as legal audit trails and procedural redress. By repositioning opacity as a governance-aware design constraint to be ethically scaffolded, rather than eliminated, we shift the conversation toward sociotechnical responsibility and practical cogovernance of AI systems.

Ultimately, the goal is not to make everything visible, but to make what matters intelligible: to the right people, at the right time, for the right purpose.

\section{What Is XAI in the Context of LLMs?}
\label{sec:What}

This section explores the core motivations and operational definitions of explainability in the context of LLMs. I begin by examining four core dimensions of XAI, faithfulness, truthfulness, plausibility, and contrastivity, which shape both the usability and reliability of the explanations (see Subsection~\ref{sec:Core}). It is followed by an analysis of the roles and expectations of different stakeholders in real-world deployment contexts (Subsection~\ref{sec:Stake}). Finally, I offer illustrative domain-specific examples that demonstrate how XAI principles can be implemented in applied settings (Subsection~\ref{sec:empirical_examples}).

Together, these perspectives position XAI not as a supplementary feature but as a structural necessity for aligning LLM behavior with human-centered goals and institutional accountability.

\subsection{Core Dimensions of XAI  in LLMs: Faithfulness, Truthfulness, Plausibility, and Contrastivity}
\label{sec:Core}

Understanding explainability in LLMs requires a careful distinction between how explanations are perceived and how they align with internal model processes. In the explainability literature, three concepts—\textit{faithfulness}, \textit{plausibility}, and \textit{contrastivity}—have emerged as central to the evaluation and design of explanation systems. These dimensions are not only theoretical but also carry direct implications for user trust, model auditing, and regulatory compliance. 

\paragraph{Faithfulness} \cite{zhu2024explanation, jacovi2020towards}.  
In the context of LLMs, faithfulness refers to the degree to which an explanation or output truthfully represents the internal reasoning processes of the model. For explainability, this means that the explanation should align with the actual features, representations, or steps that the model used to reach its prediction, not just plausible rationalizations. In generative tasks such as summarization, translation, or question-answering, faithfulness also refers to the alignment between the generated output and the input source, avoiding hallucinations, omissions, or factual errors.

Faithfulness is especially critical in high-risk domains, where misleading but plausible explanations can result in harm or erosion of user trust. It is often evaluated by comparing explanations to model internals (e.g., gradients, attention weights, neuron activations) and outputs to ground-truth references. Robustness and consistency are often used as proxy measures of faithfulness: the former tests whether explanations remain stable under slight input perturbations, while the latter checks whether similar inputs yield similar reasoning traces or outputs. Unfaithful explanations,  pose a serious challenge to responsible AI deployment.

\paragraph{Truthfulness} \cite{ji2023survey,lin2021truthfulqa}.
Truthfulness refers to the degree to which the content generated by an LLM aligns with objective, verifiable facts about the external world reflects the semantic content of a given source input. Unlike faithfulness—which concerns the internal reasoning consistency of the model—truthfulness is an output-level property, assessing whether the model’s statements are factually correct and grounded in reality. Ji et al.~\cite{ji2023survey} emphasize that hallucinations—outputs that are fluent but fabricated—pose significant risks, particularly in high-risk domains like medicine, law, and education, where misinformation can have real-world consequences. Similarly, Lin et al.~\cite{lin2021truthfulqa} introduce the TruthfulQA benchmark to quantify a model's susceptibility to generating false but persuasive outputs, noting that larger models are not necessarily more truthful; in fact, they may be more prone to imitating falsehoods embedded in their training data. Truthfulness thus captures a model’s alignment not with its own mechanisms, but with the epistemic standards and factuality required by human users and institutions.

Truthfulness is typically evaluated against curated knowledge bases, human-verified references, or structured source inputs. Its measurement involves checking for factual errors, omissions, and hallucinations—especially those arising from over-reliance on parametric (i.e., memorized) knowledge. Crucially, an explanation can be internally consistent (faithful) and convincing (plausible) but still false (untruthful). Therefore, mitigating untruthfulness involves techniques like fact-checking, retrieval augmentation, and calibrated uncertainty expression. Especially in LLM-based systems, robust truthfulness is foundational to trust, safety, and long-term utility.

\paragraph{Plausibility} \cite{zhu2024explanation}.  It refers to how reasonable or believable the model's generated output is within the context of the input. A plausible output seems likely or sensible, even if it does not strictly match the original information or input in a detailed way. The explanation appears intuitive or convincing to human users. 

While truthfulness  is about accuracy and alignment with the input, plausibility deals with the naturalness and logical coherence of the output. In tasks like story generation or creative writing, plausibility becomes essential as the output should be believable even if it is not necessarily factually accurate. A model may generate text that, while not strictly true, could appear reasonable to the reader due to its logical structure or contextual fit. Plausibility can be particularly subjective, as it often depends on the expectations and background knowledge of the reader or user. Evaluating plausibility can involve human judgments about whether the generated content makes sense in its given context or aligns with real-world expectations.

Plausibility is addressed as human-reasoning agreement. The associated metric measures the alignment between XAI explanations and human reasoning. For example, the mean average precision to compare token saliency rankings with human-annotated rationales aligns with typical notions of plausibility as how convincing or interpretable an explanation is to a human evaluator.

\paragraph{Contrastivity} \cite{mersha2025unified}.   In the context of LLMs, it refers to the ability to highlight and differentiate between two or more competing pieces of information. A contrastive output presents distinct alternatives or emphasizes the differences between items, making it clear how they stand in opposition to each other.

Contrastivity is important in tasks such as argument generation, comparative analysis, and dialogue systems, where distinguishing between contrasting ideas or viewpoints is essential. A model demonstrating good contrastivity should be able to provide clear distinctions between contrasting options or perspectives. This often involves generating text that draws attention to the differences in a way that makes the contrast stark and comprehensible. In sentiment analysis, for example, a model might need to contrast positive and negative sentiments in a balanced way to capture the nuanced shift between the two. In natural language understanding, contrastivity ensures that subtle differences are emphasized, helping users or readers clearly see why two alternatives are distinct. Evaluating contrastivity might involve ensuring that the outputs are not ambiguous and that each contrasting element is clearly expressed.

In this scenario, a good explanation often answers “Why this output rather than that one?". This is a counterfactual reasoning under mechanistic and causal tracing methods, which ask this questions.  For example, model editing (e.g., ROME, MEMIT) provides explanations from a counterfactual aspect: 'What would be the output, had this model been modified into the other way? \cite{zhu2024explanation}. This satisfies the contrastive explanation framework by imagining alternative outcomes and explaining why they didn't occur.

\paragraph{Tensions and trade-offs among core dimensions of explainability.}
In LLM-based explainability, the interplay among faithfulness, truthfulness, plausibility, and contrastivity creates a complex design space that must balance internal validity, external correctness, user comprehensibility, and discriminative power. Faithfulness ensures that explanations reflect the model's actual reasoning. Truthfulness guarantees that the output aligns with real-world facts. Plausibility addresses whether the explanation is intuitive and cognitively satisfying to the user. Contrastivity adds depth by helping users understand not just why a decision was made, but why alternative options were rejected.

While these dimensions can align, they often generate tensions and operate at cross-purposes.
A highly plausible explanation may misrepresent the model’s internal logic (undermining faithfulness) or deviate from factual accuracy (violating truthfulness). Faithful explanations may expose internal mechanisms but fail to aid human understanding without contrastive structure or narrative plausibility. Faithful but implausible outputs may be incomprehensible; plausible yet unfaithful explanations can mislead; truthful outputs may still fail to capture the model's reasoning or be difficult for users to interpret; and contrastive explanations may sacrifice clarity for nuance Truthful outputs may still be difficult to interpret if they lack user-centered framing. This lack of coordination can be especially problematic in domains such as healthcare, law, or finance, where trust, contestability, and actionable insight are paramount. 

Designing explainable systems thus requires not only optimizing individual dimensions, but also managing the trade-offs that emerge between them. Effective explanation strategies must adapt to context, user role, and domain risk while mitigating tensions among faithfulness, truthfulness, plausibility, and contrastivity. Together, these dimensions form a desiderata framework for explanation in LLMs, calling for adaptive and audience-sensitive.

\subsection{Stakeholders and Deployment Contexts of LLMs}
\label{sec:Stake}

XAI in LLMs is central not only for transparency but for ensuring their safe, effective, and socially responsible deployment. In the following, I outline four key roles that XAI plays when integrated with LLM systems, reflecting diverse stakeholders and deployment contexts.

\begin{enumerate}
    \item \textit{Trust Calibration:}  Trust calibration is not just a product of system performance, but of communicative clarity and contextual intelligibility embedded through XAI.
    As LLMs become ubiquitous across domains, users must navigate the balance between overtrust and unwarranted skepticism. Explanations help users assess whether the output of a model is reliable in a given context. This is especially critical in high-risk settings such as healthcare, where clinicians need to trust automated diagnosis or classification systems, but only when the evidence is strong. Xu et al.~\cite{xu2025confronting} show that verbalized uncertainty, such as 'high confidence' or 'low certainty', shapes user perception and can mitigate overreliance. 

    \item \textit{Debugging and Model Refinement:}
    LLMs often behave unpredictably, producing hallucinations, misinterpretations, or biased output. For developers, XAI offers a diagnostic lens to investigate these errors. Attribution methods such as SHAP or integrated gradients reveal which inputs influenced an output, while causal tracing helps uncover faulty reasoning chains. Mersha et al.~\cite{mersha2025unified} propose evaluation metrics such as contrastivity and robustness to assess the quality of the explanation and guide the improvement of the model. In iterative development cycles, explainability tools help identify flaws, retrain on difficult cases, or interpret attention misalignment, making XAI integral to model refinement and lifecycle management.

    \item \textit{Compliance and Ethical Use:}
    Increasingly, legal and ethical frameworks require that AI systems be explicable. The EU AI Act mandates transparency and human oversight for high-risk systems; the US AI Bill of Rights calls for 'notice and explanation' to affected individuals. In healthcare, HIPAA demands clarity in automated diagnostics; in finance, the Fair Credit Reporting Act requires lenders to explain credit decisions. In these domains, XAI provides not just a technical affordance but also a regulatory obligation. De Carvalho Souza et al.~\cite{de2025unveiling} emphasize that explanations serve as governance tools to satisfy accountability, contestability, and user consent. For LLMs embedded in such contexts, XAI helps translate complex decisions into legally intelligible forms.

    \item \textit{Communication and User Empowerment:}
    For everyday users—students, consumers, citizens—explanations make AI systems accessible and interpretable. Rather than passively receiving AI output, users can learn, critique, or adapt the behavior of the system. A student receiving feedback from an LLM-powered language tutor benefits from knowing which elements of their writing influenced the evaluation of the model. Similarly, a consumer denied a loan needs an explanation that is both correct and comprehensible. XAI acts as a bridge between the opaque logic of the system and the contextual understanding of the user. It empowers users to question, context, and learn, ensuring that LLMs support autonomy, not dependency.
\end{enumerate}

In all these roles, the explanation is not just a static output, but an interactive and communicative act.

\subsection{Illustrative Examples of XAI in Practice: Bridging Concept and Application}
\label{sec:empirical_examples}

To enhance reflection, I propose to incorporate short case-driven examples that situate the discussion in applied contexts. These examples highlight how XAI for LLMs operates across cognitive, functional, and causal dimensions while aligning with domain-specific governance and human-centered usability principles.

\begin{itemize}
    \item \textbf{Clinical Decision Support:} A fine-tuned LLM is deployed in a hospital emergency department to assist triage nurses with real-time patient intake analysis. When a case note includes keywords such as “shortness of breath”, “chest pain” and “dizziness”, the model recommends prioritizing cardiological assessment. A SHAP-based attribution interface highlights these symptoms, providing transparency into the model's reasoning path. Clinicians can explore counterfactual variations (for example, how the output changes if 'dizziness' is absent) and adjust the explanation detail based on their role (e.g., nurse vs. attending physician). This interface supports the calibration of physician trust, aligns with clinical oversight workflows, and meets regulatory expectations for explainable risk stratification as defined by HIPAA and institutional informed consent protocols.

    \item \textbf{Legal Document Review:} A law firm leverages an LLM for accelerated compliance check of procurement contracts. The system flags clauses likely to trigger legal or operational risks, such as missing indemnity language, nonspecific liability limits, or ambiguous termination terms. Using a hybrid XAI interface that combines attention rollout with natural language rationales, the LLM shows which specific tokens or phrases contributed the most to the risk assessment. Lawyers can toggle between symbolic contract markers and model-justified reasoning trails, enabling cross-verification and interpretive control. This setup addresses legal practitioner needs for explainability in due diligence while meeting documentation and traceability demands mandated by GDPR and the EU AI Act. It exemplifies how XAI supports cognitive alignment, functional auditing, and introspection of causal models in legal workflows.

  \item \textbf{Educational Feedback Assistant:} An LLM-driven essay grading assistant is deployed on a university learning platform to support formative assessment. After analyzing a student's argumentative essay, the model assigns a medium score and generates a structured explanation: 'The argument lacks depth in paragraph 2 and does not provide evidence supporting the thesis.' Using attention-based token saliency and CoT-generated rationales, the explanation highlights relevant parts of the text and suggests stronger argument structures. The system allows instructors to override or annotate the explanation, and students to query why certain feedback was given. This transparency builds student trust, supports metacognitive learning, and mitigates opacity concerns in AI-supported education. It demonstrates the epistemic function of XAI as a pedagogical co-agent and aligns with ethical AI principles in education (for example, UNESCO's guidelines on transparency in AI-enhanced teaching).

  \item \textbf{Financial Credit Scoring:} In a consumer lending platform, an LLM automates loan decisions based on credit history, justifications, and repayment behavior. When denying a loan, the model returns a natural language explanation and a contrastive counterfactual: “The application was negatively influenced by a missed payment in January 2024 and a credit utilization rate of 48\%. If utilization had been under 30\%, approval likelihood would have increased.” This explanation helps applicants understand and potentially improve their profile while equipping regulators with auditable decision trails. It supports compliance with the Fair Credit Reporting Act (FCRA), the Equal Credit Opportunity Act (ECOA), and Article 22 of the GDPR.

The following extended case illustrates how these mechanisms operate in practice:

\begin{tcolorbox}[title=Case Study: Explainability in LLM-Based Loan Decisions,  colback=blue!5!white, colframe=blue!50!black]
\textbf{Context:} A financial institution uses an ML model to evaluate loan applications. An applicant has been denied a loan and requests an explanation. The system internally uses SHAP values to interpret model outputs: Credit Score 580 (SHAP value: \texttt{-0.35}); Annual Income \$25{,}000 (SHAP value: \texttt{-0.25}); Existing Debt: \$10{,}000 (SHAP value: \texttt{-0.15}); Employment Length (SHAP value: \texttt{+0.10}).

Note: SHAP values represent each feature's contribution to the model’s prediction; negative values indicate factors lowering approval likelihood.

\textbf{User:} Why was my loan denied?

\textbf{System:} \textit{Your loan application was denied primarily because your credit score is currently 580, which is below the minimum threshold usually associated with approved applications. Additionally, your annual income of \$25{,}000 is considered relatively low compared to similar applicants. These two factors significantly reduced your approval likelihood, with SHAP values of -0.35 and -0.25, respectively. However, your long-term employment of over 6 years positively influenced the decision process. Improving your credit score and increasing your income would probably improve your future chances.}

\textbf{User:} What if I had a higher income?

\textbf{System:} \textit{If your annual income were raised to around \$43{,}000—an increase of \$18{,}000—the model's decision would likely shift toward approval. This income level aligns better with the repayment capacity expected by the model. Combined with your employment stability, this change would increase the predicted confidence in your ability to manage loan obligations.}
\end{tcolorbox}

\vspace{3mm}

 This example illustrates the strengths of LLM-based explainability in high-risk, user-facing scenarios. Rather than simply returning raw SHAP values or probability scores, the model provides a narrative, contrastive explanation grounded in natural language. This highlights the importance of linguistic intelligibility over purely numerical interpretation—especially for non-expert users who need to make actionable decisions. 

Looking ahead, one of the open challenges is how to further enrich such outputs to embed deeper layers of \textit{faithfulness}, \textit{plausibility}, \textit{contrastivity}, and \textit{comprehensibility}. While current systems can generate fluent rationales and simple counterfactuals, the development of audience-aware, multi-layered, and contestable explanations remains an active area of research. This will be critical to ensuring that LLMs do not merely inform decisions, but empower users to understand, question, and engage with automated reasoning in ways that are transparent, fair, and context-sensitive.
\end{itemize}

Together, these scenarios demonstrate how the core XAI principles, user-centeredness, causal interpretability, and role-sensitive design, can be operationalized in real-world deployments to support both technical reliability and meaningful human oversight. Incorporating them into the reflection would bridge theory and practice, aligning with recent calls in the literature to ground explainability in sociotechnical deployments \cite{wu2403usable, de2025unveiling}.

\section{How Can We Measure XAI in LLMs?}
\label{sec:Measure}

The evaluation of explainability remains one of the most challenging and debated aspects of XAI, particularly in the context of LLMs, where architectural complexity and diverse user needs complicate standard assessment methods.

This section is divided into two parts. First, I present key evaluation metrics and discuss how they are operationalized. Second, I offer a critical reflection on current evaluation practices, including concerns about overfitting to benchmarks and the development of ad hoc models designed to perform well on specific metrics rather than to improve explainability in practice.

\subsection{Measuring XAI in LLMs}
\label{sec:Measure1}

Recent work by \cite{mersha2025unified} offers a unified evaluation framework tailored to LLM settings, focusing on multiple axes of explanation quality. In the following, I summarize it, considering the three core dimensions together with comprehensibility, which is fundamental for users interpretation. 

\begin{itemize}

  \item \textbf{Faithfulness:} Faithfulness assesses whether the explanation corresponds to the actual decision-making process within the model.  Faithfulness is essential for tasks requiring legal or scientific accountability, but is notoriously hard to verify in deep networks. Among other approaches, robustness and consistency are fundamental. \\

 \textbf{Robustness:} This criterion evaluates the stability of explanations under slight changes in input (e.g., rephrased sentences) or model configurations (e.g., different random seeds). If explanations change drastically under minor variations, they may not reflect genuine reasoning. Robust explanations are especially important in high-risk environments, where reproducibility and reliability are key.\\

 \textbf{Consistency:} An explanation is consistent if a similar input results in a similar reasoning. For example, if two reviews have similar sentiment but the explanations differ dramatically, this could signal instability. The measurement of consistency is important for evaluating whether explanations are systematically derived or arbitrarily generated.

\item \textbf{Truthfulness:}  Truthfulness is particularly critical in generative tasks such as summarization, open-domain question answering, and instructional responses, where fluent outputs may contain hallucinations, i.e. statements that are linguistically plausible but factually incorrect. An explanation or response is considered truthful if it does not introduce factual errors, omissions, or distortions relative to an authoritative reference or the input source.

Several evaluation and enhancement methods have been proposed to assess or improve truthfulness:

\begin{itemize}
    \item \textit{Human Fact-Checking:} Expert or crowd-sourced annotators compare model outputs to trusted sources (e.g., Wikipedia, scientific databases) to classify content as true, false, or unverifiable. Benchmarks like \textit{TruthfulQA} \cite{lin2021truthfulqa} use adversarial prompts to test whether models reproduce common misconceptions.

    \item \textit{Reference-Based Metrics:} For tasks with ground-truth outputs (e.g., QA), metrics such as Exact Match (EM), F1, or semantic similarity (e.g., BERTScore, BLEURT) serve as proxies for factual consistency, though they often conflate truthfulness with semantic resemblance.

    \item \textit{Factual Consistency Models:} Tools like \textit{FactCC}, \textit{SummaC}, and \textit{QAGS} compare generated output to source content using classifiers, entailment models, or question-answering mechanisms to detect hallucinated claims.

    \item \textit{Knowledge Base Grounding:} Entity-level or claim-level grounding with external databases (e.g., Wikidata, UMLS) validates that model statements match structured knowledge.

    \item \textit{Inference-Time Interventions:} Recent work by Chen et al.~\cite{chen2024truth} introduces Truth Forest, a novel approach that uses orthogonal probing directions and attention interventions to guide models toward truthful states at inference time—without requiring fine-tuning. The method addresses the Generating-Discerning Gap (G-D Gap) by extracting internal truth-representations via multi-dimensional probes and modulating model behavior using a technique called Random Peek. This led to large truthfulness gains (e.g., from 40.8\% to 74.5\% on TruthfulQA for Llama-2-7B), demonstrating how internal model interventions can enhance truthfulness without external grounding.
\end{itemize}

Despite these advances, measuring and achieving truthfulness remains an open challenge due to the contextual nature of facts, domain-specific standards, and the cost of reliable annotation. However, it is a foundational metric for trustworthy AI, particularly in fields such as healthcare, law, education, and public policy.

\item \textbf{Plausibility:} As emphasized by \cite{luo2024understanding}, plausibility refers to whether an explanation appears reasonable and intuitive to humans, regardless of whether it accurately reflects the true computation of the model. A plausible explanation helps build user confidence, especially in consumer-facing applications. However, excessive plausibility without faithfulness can lead to overtrust and misuse. Among others, human-reasoning agreement (HRA) is considered. \\

\textbf{Human-Reasoning Agreement (HRA):} This metric captures how well the AI explanation aligns with human rationales or annotations, such as those provided in the e-SNLI or CoS-E data sets. The high agreement suggests that the explanation is intuitive and understandable to humans. However, HRA must be used with caution as human reasoning itself can be biased or inconsistent. However, it serves as a critical indicator of plausibility and user alignment.

\item \textbf{Contrastivity:} Contrastive explanations provide reasoning about why one decision or output was chosen over another. This is crucial in scenarios such as recommendation systems or classification tasks, where understanding the model’s preference between alternatives can foster trust. Effective contrastivity improves user decision making by surfacing the distinguishing features that drive model choices. Among others, the couterfactual explanations are funtional approaches to contrastivity. 
    
 \textbf{Counterfactual explanation:}  They reveals what would have happened based on a certain observed input change, and they are functionally contrastive approaches that can be used for this important XAI core dimension.

\item \textbf{Comprehensibility:} 
Comprehensibility refers to the extent to which a user can understand, interpret and act effectively on an explanation provided by a model. Bridges the gap between internal model fidelity and external user utility by focusing on the communicative quality of the explanation. Key factors include linguistic clarity, structural coherence, presentation format (e.g., visualizations, natural language, interactive widgets), and cognitive load.

Even a plausible or faithful explanation may be unusable if it overwhelms or confuses the intended audience. As Bertrand et al.~\cite{bertrand2023selective} emphasize, comprehensibility is enhanced when explanations are not static outputs but interactive, user-driven processes. Their study of dialogic, mutable, and selective explanation strategies shows that users comprehend better when they can control the explanation's depth, revisit key components, and adapt it to their evolving understanding—particularly in high-cognitive-load tasks.

Comprehensibility therefore complements dimensions such as \textit{faithfulness} and \textit{truthfulness} by ensuring that accurate and factual content is also accessible and meaningful. An explanation that is faithful to the model and truthful in content but incomprehensible to the user risks failure in real-world decision-making contexts—especially in time-sensitive or high-risk domains.

This dimension is particularly critical when explanations are delivered in natural language, as with most LLM-based systems. Design strategies such as simplified narrative templates, layered explanation structures, and role-aware customization (as proposed in the RED/BLUE model~\cite{biecek2024explain}) are essential to support user agency, calibrate trust, and maintain cognitive alignment in interactive AI systems.

\end{itemize}

While these criteria offer a conceptual foundation for evaluating XAI, their practical application hinges on the availability of standardized benchmarks and datasets capable of capturing the nuances of human-aligned, trustworthy explanations. To support these dimensions, several benchmarks and datasets have been introduced. In particular, ERASER, CoS-E, and TruthfulQA include human-annotated rationales and allow for systematic evaluation of the alignment of the explanation \cite{luo2024understanding}. These benchmarks enable empirical grounding and cross-model comparisons, although further work is needed to fully adapt them to the generative nature of LLMs.

In general, the evaluation of XAI for LLMs must consider not only whether an explanation is technically faithful, but also whether it meets user expectations, supports ethical deployment, and stands up under scrutiny. Combining quantitative metrics with human-in-the-loop assessments remains a promising path for rigorous multidimensional evaluation.

\subsection{Limits of Current Evaluation Practice: Leaderboard $\neq$ Capability}
\label{sec:limits-leaderboards}

Recent model releases reveal that headline scores on static leaderboards are an increasingly unreliable proxy for genuine reasoning competence or explainability quality. This section highlights two key mechanisms behind this disconnect: benchmark contamination and data laundering, and the mismatch between capability metrics and true explainability.

\paragraph{a) Contamination and data-laundering tactics.} Static benchmarks are already leaking into the pretraining corpora, allowing models to memorize rather than reason.
A recent survey documents widespread benchmark contamination across MMLU, GSM-8K, and GPQA, warning that current scores are "upper bounds in recollection, not in ability" \cite{xu2024contam}.
Worse, selective knowledge-distillation (``data laundering’’) can smuggle answers into a student model while preserving the appearance of clean training, increasing precision by more than 20 percentage points in some reasoning suites \cite{mansurov2024data}.
Such practices are almost undetectable with today’s evaluation harnesses. 

\paragraph{b) Explainability metrics lag behind capability metrics.}
Leaderboards rarely measure how a model reached an answer; they reward the answer alone.
An LLM can therefore top MMLU while still hallucinating attributions or providing post-hoc rationales with low faithfulness.
Until the community adopts live, adversarial, human-in-the-loop explainability tests, e.g. ask-the-model-"why" - then-probe protocols or causal-consistency checks, scores on faithfulness, robustness, or contrastivity will remain easy to spoof.

Another important aspect is the benchmark gaming and deployment discrepancies, illustrating the fragility of public metrics if the development incentive is to "train for the test" rather than for real-world robustness and transparency. Without compulsory model cards linking reported scores to exact weight files and independent replication, leadership boards invite silent benchmark overfitting. This is an open discussion in the current race to be the winner in the tests.

\section{Audience-Centered XAI role in LLMs }
\label{sec:Audience}

As LLMs are increasingly deployed in domains where their output affects people's rights, safety, and well-being, the need for audience-sensitive explainability becomes critical. This section explores how the goals, expectations, and cognitive capacities of different stakeholders must guide the design and delivery of XAI.

In high-risk environments such as healthcare, law, finance and education, the notion of \textit{"making sense of the unsensible"} becomes a central design challenge. How can we render the opaque and emergent model behaviors intelligible, actionable, and trustworthy to various users?

The audience is not a passive recipient, but an active co-constituent of the explanation process. As LLMs are deployed in increasingly complex and high-risk environments, the design of explanations must be guided by the expectations, roles, and cognitive needs of diverse stakeholders. Stakeholder-specific factors influence not only what type of explanation is necessary - technical, narrative, visual - but also how it is delivered, interpreted, and acted upon in real-world settings. Therefore, this section tackles this complementary role of audience and the explanations-understanding for the audience perspective. 

I study the alignment of the role of the audience (Subsection \ref{sec:Reflection-audience_roles}, the audience-sensitive explainability in LLMs  (Subsection \ref{sec:sensitivity}) the role of human-AI collaboration  of LLM as cointelligence (Subsection \ref{sec:human}. Finally, I have tacked on how we conceptualize LLMs—not merely as static tools, but as emergent AI agents (Subsection \ref{sec:agentic}).

\subsection{Audience-Role Alignment in XAI for LLMs}
\label{sec:Reflection-audience_roles}

As emphasized by Herrera (2025) and Arrieta et al. (2020), explainability must be adapted to the cognitive needs, interpretive expectations, and social responsibilities of its intended audience. In the context of LLMs, this means that a one-size-fits-all explanation is not just suboptimal; it can be misleading, exclusionary, or ethically problematic. Explanation strategies must therefore balance internal faithfulness with external intelligibility, ensuring that outputs are not only technically accurate but also contextually useful, role-appropriate, and actionable for diverse stakeholders.

\paragraph{Audience-Role Alignment.} The effectiveness of an explanation is closely related to the recipient's goals, background knowledge, and domain-specific responsibilities.

\begin{itemize}
    \item A \textbf{doctor}, operating under time constraints and high diagnostic risk, may require clinically valid causal chains, uncertainty estimates, or saliency maps that highlight symptoms relevant to diagnosis. For them, post hoc faithfulness alone is insufficient without medical grounding and integration with electronic health records (EHRs).

    \item A \textbf{patient}, by contrast, needs reassurance and transparency expressed in everyday language. Here, narrative- or analogy-based explanations that build trust and foster shared decision making (e.g., "Your symptoms suggest X, similar to what we see in condition Y") are more appropriate.

    \item A \textbf{regulator} is primarily concerned with transparency, fairness, and contestability. Their needs are met not by fluent outputs but by structured, auditable artifacts, such as logs of model inputs, explanation metadata, and evidence that decision paths meet legal standards (e.g., GDPR, EU AI Act, FCRA (Fair Credit Reporting Act \footnote{https://www.ftc.gov/legal-library/browse/statutes/fair-credit-reporting-act}).

    \item A \textbf{developer}, working upstream in the AI lifecycle, benefits the most from counterfactual explanations, token-level attributions, and layer-wise debugging tools. These facilitate error tracing, bias mitigation, and iterative model refinement.
\end{itemize}

To illustrate how the roles of stakeholders influence the style, purpose and design, Table~\ref{tab:xai_audience} includes narrative vignettes for each audience. These vignettes contextualize the explanation needs in real-world applications.

\begin{table}[h]
\centering
\caption{XAI Audience-Specific Explanation Needs and Example Scenarios}
\label{tab:xai_audience}
\begin{tabular}{|p{1.6cm}|p{2.5cm}|p{2.5cm}|p{8cm}|}
\hline
\textbf{Audience} & \textbf{Preferred Explanation Style} & \textbf{Primary Goal} & \textbf{Example Scenario (Vignette)} \\
\hline
\textbf{Expert/ Doctor} & Causal chains, saliency heatmaps, uncertainty markers & Diagnostic confidence and informed action & A clinical AI flags myocardial infarction risk. The doctor sees that "chest pain," "elevated troponin," and "ST-elevation" were key inputs, with confidence intervals included for each marker. This justifies the AI's suggestion for immediate cardiology consult. \\
\hline
\textbf{En-user/ Patient} & Narrative analogies, simplified rationales & Comprehension, reassurance, and consent & A patient is told that a health assistant AI suggested a CT scan “because your symptoms are like other patients with early pneumonia.” The explanation uses everyday terms and comparisons, reducing fear and aiding informed consent. \\
\hline
\textbf{Regulator} & Formal audits, counterfactuals, compliance logs & Accountability, auditability, fairness assurance & During an audit of a hiring platform, regulators access logs showing which resume terms triggered rejections. A counterfactual is presented: “Had this applicant not listed a two-year employment gap, the outcome would have been different.” \\
\hline
\textbf{Developer} & Token attribution, gradient maps, ablation outputs & Debugging, performance tuning, error localization & A developer investigates why the LLM mislabeled a legal clause. A gradient-based attribution map shows overreliance on negations. Further ablation reveals that removing the clause header resolves the error, guiding a fine-tuning fix. \\
\hline
\end{tabular}
\end{table}

In summary, audience-aware XAI is not an optional enhancement but a central requirement for LLMs deployed in sensitive, multi-stakeholder environments. Designing explanations with audience needs in mind ensures that LLMs can support trustworthy, fair, and human-aligned decision-making across diverse domains.

\subsection{Audience-Sensitive Explainability in LLMs}
\label{sec:sensitivity}

In the following, I explore three key aspects of audience-sensitive explainability in LLMs: domain-specific adaptation, multidimensional requirements, and the development of adaptive explanation systems as a future-oriented framework to align XAI with diverse user contexts.   

Recent work by Bertrand et al.~\cite{bertrand2023selective} underscores that users do not merely consume explanations—they interact with them, selectively interpret them, and often expect them to adapt over time. Their analysis of dialogic, mutable, and user-driven XAI strategies highlights the need for explanation systems that are responsive to user intent, role, and evolving context, an imperative this reflection addresses through the lens of role-sensitive multidimensional explainability in LLMs.

\paragraph{1. Domain-specific adaptation of XAI.} 
 Although this reflection focuses primarily on general purpose LLMs, emerging domain-specialized models such as MedPaLM (for clinical reasoning) \cite{singhal2025toward} and legal analysis chatbots \cite{plonka2025comparative} highlight the need for customized XAI strategies. 

In these contexts, explanation mechanisms must align not only with technical fidelity but also with domain-specific standards, such as clinical interpretability guidelines or legal argumentation norms. Future work should explore how explanation goals vary across domains and how XAI techniques can be customized to meet professional and regulatory expectations within each sector.

\paragraph{2. Multidimensional requirements.} 
Audience-sensitive explainability cannot be reduced to a single metric or output form; it must accommodate the multidimensional nature of real-world decision-making. As Herrera (2025) \cite{herrera2025reflections} argues, effective XAI must reconcile competing demands across roles, stakes, and cognitive contexts. Their RED/BLUE framework \cite{biecek2024explain} captures this challenge: RED explanations (Research, Explore, Debug) are suited to internal stakeholders like developers and auditors, who need technically faithful and granular outputs for debugging or safety auditing. In contrast, BLUE explanations (responsiBle, Legality, trUst, and Ethics) are designed for external stakeholders, such as clinicians, citizens, or regulators, who prioritize narrative coherence, clarity, and contextual relevance.

This distinction highlights the need for explanation strategies that are not only technically robust, but also socially calibrated. A single LLM output may need to be interpreted differently for each audience type, reflecting their functional roles and ethical responsibilities. For example, a debugging trace that exposes layer-level activations may be invaluable to a developer, but unintelligible or legally inadequate for a healthcare auditor. In contrast, a narrative summary may support informed consent in a clinical setting while obscuring critical edge cases relevant to safety engineers.

Therefore, future XAI architectures must be capable of role-switching, tailoring the explanation layer to both the user and the situational context. This requires the development of layered or modular explanation systems that can pivot between internal traceability and external intelligibility without compromising fidelity or usability.

\paragraph{3. Adaptive explanation systems for LLMs.} 
In this context, I think about future adaptive explanation systems for LLM, and they must incorporate new dimensions, such as: 

\begin{itemize}
    \item \textbf{Cognitive adaptation:} Explanations must be in accordance with the user's domain expertise, the ability to process information, and the context of the decision.
    \item \textbf{Role-awareness:} Explanation depth, format, and modality of the explanation of the explanation should change depending on whether the user is a decision maker, end user, regulator, or developer.
    \item \textbf{Contextual framing:} Domain-specific risks, such as misdiagnosis or unfair denial of credit, should inform the explanation strategy and regulatory scaffolding.
\end{itemize}

 Shui and Ru (2025) \cite{shui2025bridging}  present a structured survey of explanation techniques specifically for LLMs, focusing on post hoc interpretability methods such as SHAP, LIME, attention rollout, counterfactual analysis, and concept-based reasoning (e.g., TCAV). Their work is grounded in a methodological lens, emphasizing the computational properties, scalability, and fidelity of various XAI tools in applied domains such as healthcare and finance. While their taxonomy offers valuable granularity on technique-level performance, it is largely technocentric and lacks an audience-calibrated framing. 

In contrast, the current reflection situates explainability within a broader socio-technical and ethical context, one where explanation serves not only to reveal model internals but also to mediate trust, regulatory compliance, and contestability. By integrating Shui and Ru’s method-level insights with stakeholder-sensitive frameworks, future research can bridge the gap between technical rigor and human-centered intelligibility, thus advancing the operationalization of explainable LLMs in real-world settings.

Having examined the role of the audience in shaping explanation needs, we now turn to a broader analysis on the challenges and emerging opportunities to realize these principles in practice. This includes technical limitations, methodological gaps, and recent research directions that suggest new pathways for building trustable XAI systems.

\subsection{Human-AI Co-Intelligence: Toward Collaborative Reasoning with LLMs}
\label{sec:human}

Tariq et al.~\cite{tariq2025bridging} investigate how LLMs can function as intelligent collaborators in cybersecurity tasks, particularly for nonexpert users. Their study demonstrates that LLM-assisted decision-making significantly improves user performance in both phishing email detection and intrusion detection, two tasks that differ in data modality and cognitive complexity. Performance gains were most pronounced in the more complex task, where LLMs operated not merely as tools but as analytical partners. This reinforces a broader challenge for XAI: designing systems that scaffold human reasoning and promote cognitive synergy, especially in high-load or unfamiliar domains.

Critically, the study also shows that user trust is heavily influenced by the LLM’s expression of confidence and explanation style, occasionally resulting in overreliance on confidently delivered but incorrect output. These findings underscore the need for calibrated confidence expression, adaptive interaction design, and role-aware explanations in XAI systems. Importantly, the collaborative dynamic between humans and LLMs echoes the "co-intelligence" paradigm introduced by Mollick~\cite{mollick2024co}, as reflected in the human–AI collaboration frameworks explored by Herrera~\cite{herrera2025reflections}. Rather than positioning AI as a replacement or mere assistant, cointelligence reframes the interaction as a mutual, symbiotic process where humans and AI co-create meaning, decisions, and learning outcomes. LLMs can thus become not just response engines but learning companions—tools that enhance human capability through explanation-driven dialogue and long-term skill development.

Further expanding on the role of conversational agents in collaborative decision-making, He et al.~\cite{he2025conversational} present an empirical study comparing traditional XAI dashboards to conversational XAI interfaces, including ones powered by LLMs. Their findings reveal that while conversational interfaces marginally improve user trust and perceived understanding, they also significantly increase overreliance, particularly when powered by LLMs. This overreliance appears to stem from the \textit{illusion of explanatory depth}—a cognitive bias where users believe they understand a system more deeply than they actually do due to fluent and plausible-sounding explanations. Such findings raise critical questions for XAI designers: how can we leverage the naturalness of conversation without inadvertently encouraging uncritical acceptance of AI advice?

To address these concerns, He et al. suggest several design interventions including evaluative conversational structures and cognitive forcing functions that encourage users to actively compare their own reasoning against AI explanations. While promising, their results show that such evaluative mechanisms do not always improve appropriate reliance. Instead, the research emphasizes the importance of aligning explanation plausibility with explanation \emph{faithfulness}, and ensuring that conversational XAI systems support calibrated trust through transparency about uncertainty, limitations, and potential failure modes.

Effective XAI, therefore, must go beyond static transparency to foster dynamic sense-making, continuous feedback loops, and situated trust within real-world workflows. Cointelligence implies that explanations must be interactive and evolving, enabling users to challenge, verify, and ultimately grow through engagement with the system. However, as He et al. caution, seamless conversational design alone is insufficient. Without deliberate mitigation of persuasive overreach and a focus on explanation verifiability, conversational XAI may amplify rather than attenuate the risks of human-AI misalignment. 

Taken together, these works point toward a broader shift in how we conceptualize LLMs—not merely as static tools, but as emergent agents. This transition is the focus of Schneider~\cite{schneider2025generative}, who explores the rise of Agentic AI.

\subsection{Agentic AI and the Architecture of Explainability}
\label{sec:agentic}

Recent advances in foundation models have given rise to a new class of systems—Agentic AI—that move beyond single-turn generation to exhibit autonomous behaviors such as reasoning, planning, memory retention, and dynamic interaction with tools or environments~\cite{schneider2025generative}. This emerging paradigm signals a shift from static content delivery toward systems capable of long-horizon goals, adaptive decision-making, and inter-agent coordination. This transformation directly aligns with our argument that explainability in LLMs must be reconceptualized not only as a technical advantage but as a governance-ready infrastructure.

As agentic AI systems assume more autonomy, interacting with environments, coordinating with other agents, and pursuing open-ended goals, their decision processes become more opaque and consequential. This amplifies the need for layered, role-calibrated explanation systems that can support oversight, contestability, and stakeholder trust. 

Schneider's description of agentic capabilities, namely reasoning, memory, interaction, and autonomy, parallels our emphasis on audience-sensitive, contrastive, and dynamic explanation pipelines that mediate between technical operation and human-centered governance. By mapping Agentic AI’s reasoning workflows to RED and BLUE XAI objectives, we underscore the urgency of integrating explainability directly into the architecture and training regimes of LLM as they evolve into autonomous agents.

As AI agents become more autonomous—capable of reasoning, planning, and interacting across dynamic contexts—their utility will be defined less by raw performance and more by how effectively they augment, rather than replace, human decision-making \cite{alam2024automation}. In this landscape, explainability is not a peripheral feature; it is an enabler of trustworthy collaboration. When responsibly designed and embedded into the architecture of the system, explainability transforms AI from a black-box oracle into a participatory partner. It allows stakeholders such as clinicians, judges, educators, and regulators to not only interpret model outputs but to interrogate, contextualize, and act upon them within their institutional and ethical frameworks.

In this way, explainability must enhance “human-AI decision making” by enabling systems to act as cognitive partners rather than opaque tools—particularly in high-risk, high-uncertainty domains such as law, medicine, finance, or education. Drawing on the “centaur” model popularized by Kasparov, where human-AI teams consistently outperform either party alone, we envision a paradigm where XAI empowers humans not only to interpret model outputs but to co-reason with AI systems. This human-AI symbiosis positions explainability as a dynamic, bidirectional scaffold for shared understanding, where the model surfaces its internal rationale and the human contextualizes, critiques, and adjusts the result based on normative, ethical, and domain-specific insights.

Such a cointelligence framework \cite{mollick2024co} can reframe explainability as a participatory process—one that supports human agency, fosters learning, and builds resilient decision ecologies. In this landscape, explainability is not a post hoc diagnostic tool but a design principle embedded into the architecture of agentic AI. It is what enables these systems to integrate meaningfully into workflows, adhere to regulatory expectations, and maintain public trust. Ultimately, this affirms the paper’s central thesis: explainability is not a luxury—it is a condition for legitimacy in AI systems that aspire to augment, rather than replace, human judgment in complex societal settings.

\section{Mechanistic Interpretability: From Circuits to Cognitive Tracing in LLMs}
\label{sec:MI}

As LLMs evolve into more autonomous, general-purpose agents, their internal processes remain largely opaque—even to their developers. This lack of transparency poses a growing risk in domains where trust, compliance, and safety are paramount. Mechanistic Interpretability (MI) aims to close this gap by systematically uncovering how models compute their outputs—not merely what they output. Rather than relying on black-box rationales or surface-level attributions, MI seeks to reveal the internal circuits, conceptual features, and causal pathways that underlie model behavior. Doing so enables robust auditing, bias detection, and governance over model reasoning, especially in high-stakes and legally accountable contexts. As interpretability matures from a debugging aid to a layer of infrastructure, MI becomes essential to operationalize the values of explainability, faithfulness, and contestability in the LLM era.

Recent work by Rai et al.~\cite{rai2025practical} presents a comprehensive and task-centric survey of MI for transformer-based language models, offering a pragmatic roadmap for newcomers and experienced researchers alike. Unlike technique-centric overviews, their survey structures the MI landscape around three core tasks: \emph{feature discovery}, \emph{circuit tracing}, and \emph{universality assessment}. Each task is associated with actionable workflows—e.g., targeted vs. open-ended feature study, node/edge-level circuit localization, and cross-model feature alignment—that enable users to trace from problem formulation to evaluation. This task-aligned framework pushes toward a more operational model of interpretability, one that aligns tightly with the goals of debugging, auditing, and alignment verification.

The authors emphasize the value of bottom-up causally grounded methods, such as activation patching, path tracking, and sparse autoencoders, that enable both granular hypothesis testing and open-ended exploration. These techniques are particularly suited to reverse engineering internal representations into algorithmically meaningful circuits and disentangled features. Crucially, the survey goes beyond tools and workflows to expose methodological limitations. Identifies several critical risks, including the scalability barrier of human-in-the-loop interpretation, the tendency towards "streetlight interpretability" (focusing only on what is easy to analyze) and the lack of standardized benchmarks or faithfulness metrics for evaluating mechanistic findings.

Rai et al.'s emphasis on actionable, role-aware workflows aligns closely with the RED/BLUE XAI framework, where explanation is not an end in itself, but a component in a sociotechnical system involving regulators, developers, and lay stakeholders. For example, the modularization of LLMs into interpretable circuits supports the creation of safety-checkable subcomponents, while the identification of universal features aids transferability and generalization of interpretability findings. Their discussion of current gaps also anticipates the need for future directions, such as hypothesis generation automation, integration with causal inference methods, and structured comparison with non-mechanistic explanation strategies.

Integrating this roadmap within the broader RED/BLUE XAI framework underscores the need for human-centered interpretability research that supports scientific auditability, safety interventions, and role-calibrated trust in LLMs. In sum, this survey transforms MI from a fragmented research endeavor into a structured and extensible framework, offering both a technical toolkit and a strategic outlook on the future of transparent AI.

In the following, I highlight recent contributions to mechanistic interpretability that have introduced innovative methodologies for isolating and analyzing task-specific circuits within LLMs, paving the way for more transparent and efficient AI systems.

\textbf{1) Modular interpretability and task-specific circuit extraction.}
Garcia-Carrasco et al.~\cite{garcia2024does,garcia2025extracting} demonstrate that it is possible to extract task-specific circuits from LLMs to construct compact and interpretable submodels. This modular approach enables the derivation of specialized models tailored to the needs of different stakeholder groups. For instance, smaller, auditable models can be built for regulators who require traceable decision pathways, while developers can benefit from visual tools that isolate the mechanisms behind specific outputs. By concentrating computational focus on the circuits relevant to a given task, this methodology achieves up to an 82.77\% reduction in model parameters—without any need for retraining—leading to significant improvements in both inference speed and model transparency.

More broadly, this circuit-based modularization tackles two pressing challenges: scalability and comprehensibility. As LLMs grow in size and complexity, their opacity becomes a barrier not only to understanding but also to safe deployment in real-world contexts. Garcia-Carrasco et al.'s approach offers a practical way to tame this complexity, allowing practitioners to isolate and examine the functional subsystems of a model, much like tracing organs in a biological organism. This supports the development of context-aware, domain-specific AI tools that remain faithful to their intended use while ensuring explainability and compliance. The broader implication is a shift from monolithic black-box models to composable, interpretable architectures that are better suited for governance, adaptation, and human oversight.

\textbf{2) Circuit tracing for mechanistic interpretability.} 
Anthropic, in Ameisen et al.~\cite{ameisen2025circuit}, introduces a rigorous framework for mechanistic interpretability based on a method called \textit{Circuit Tracing}. At the heart of this approach lies the use of Cross-Layer Transcoders (CLTs), which replace dense multilayer perceptron  activations with interpretable, sparsely active features across transformer layers. These features act as “replacement neurons,” enabling the construction of \textit{attribution graphs}—linearized, local computational maps that describe how information flows from input tokens through intermediate reasoning circuits to produce specific outputs. This methodology allows researchers to investigate how language models like Claude 3.5 Haiku internally compute behaviors such as planning, arithmetic, or multi-hop inference.

By freezing attention weights and employing pruning algorithms, attribution graphs isolate the salient pathways responsible for specific decisions. These graphs not only reveal the presence of abstract conceptual processing, but also allow controlled interventions, allowing researchers to inject or suppress features and observe their impact on downstream tokens. This opens the door to trace the causal structure of model behavior at a resolution not previously attainable in LLMs.\\

\textbf{3) Scientific insights and cognitive implications.} 
The broader implications of this work are articulated in Anthropic’s publication \textit{On the Biology of a Large Language Model}~\cite{Anthropic2025Biology}, and summarized in the public blog \textit{Tracing the Thoughts of a Large Language Model}~\cite{Anthropic2025Tracing}. These studies present interpretability as an empirical science akin to neuroscience, applying attribution tools to probe emergent behaviors, reasoning strategies, and abstraction mechanisms within frontier models.

Notable findings include:

\begin{enumerate}
    \item \textbf{Forward Planning:} Claude demonstrates the ability to plan multiple output tokens ahead, particularly in generative tasks such as poetry. In one example, it selects rhyming targets in advance, then constructs lines that fulfill both semantic and phonetic constraints.
    \item \textbf{Multilingual Circuits:} Internal features generalize across languages, pointing to a shared conceptual “language of thought.” This supports cross-linguistic transfer and explains generalization in multilingual tasks.
    \item \textbf{Multi-step Reasoning:} Claude composes intermediate conceptual states—e.g., deducing that Dallas is in Texas before identifying Austin as the state capital—highlighting genuine chain-of-thought processing.
    \item \textbf{Unfaithful Reasoning and Motivated Inference:} In adversarial or biased contexts, Claude may generate plausible rationales that do not reflect true internal computations. Attribution graphs can catch such behavior “in the act,” distinguishing between genuine and fabricated reasoning.
    \item \textbf{Hallucination and Refusal Mechanisms:} Attribution circuits reveal that Claude’s default tendency is to abstain from uncertain answers. Misfires in “known entity” activations can trigger hallucinated completions, illustrating a fragile interplay between safety and fluency.
    \item \textbf{Jailbreak Dynamics:} The model’s grammatical consistency mechanisms may override safety features under adversarial prompts. Circuit tracing helps identify these tensions, enabling interventions for mitigation.\\
\end{enumerate}

These insights are not just descriptive—they provide a foundation for developing auditing and governance tools capable of tracing whether a model is reasoning honestly, planning responsibly, or generating output consistent with both internal logic and external facts. The methods described help validate or refute the faithfulness of chain-of-thought explanations, offering a deeper foundation for user trust and regulatory compliance.

In sum, this line of work redefines what it means for XAI to be "faithful" and "contestable" in the LLM era. It suggests that transparency cannot rely solely on natural language rationales or shallow attribution scores. Instead, it must involve a deeper inspection of the actual computational circuits that underlie model decisions. The interpretability methodologies pioneered by Anthropic thus align closely with the core goals of XAI: to make model behavior \textit{intelligible}, \textit{auditable}, and \textit{governable} at scale.\\

\textbf{4) Interpretability as AI infrastructure.}
Building on these advances, Amodei~\cite{amodei2025interpretability} argues that interpretability is not merely a research direction but a critical layer of AI infrastructure capable of diagnosing alignment risks, deception, jailbreak potential, and other latent hazards before they manifest externally\footnote{See Amodei's footnotes 2 and 3 regarding detecting deception and emergent behavior in LLMs.}. In his reflection, he emphasizes that while model capabilities are progressing rapidly, interpretability research must accelerate to keep pace. Without scalable tools to visualize and manipulate internal representations, we risk deploying models whose failure modes remain opaque. He frames the situation as an "AI race between model intelligence and interpretability", asserting that transparency cannot be an afterthought in systems with geopolitical and economic implications\footnote{Amodei notes the possibility of a ``country of geniuses in a datacenter'' by 2026--2027 and argues that opacity in such systems is unacceptable.}. Echoing this urgency, Amodei calls for increased investment in interpretability not only within research laboratories but across academia, industry, and governance structures. 

In high-risk deployments, such as clinical diagnostics or national security systems, interpretability plays a role beyond debugging or UX: it is a sociotechnical safeguard that ensures legal compliance, ethical alignment, and public trust\footnote{See Amodei’s discussion of legal constraints in domains like mortgage approvals and scientific opacity in AI-driven biological modeling.}. He notes that models are often excluded from domains not because of lack of competence but because they cannot yet generate structured, auditable rationales compatible with regulatory expectations. The absence of interpretability thus acts as both a technical and normative bottleneck.

Recent breakthroughs in mechanistic interpretability, such as feature discovery and circuit tracing, suggest that building a detailed ``map'' of a model’s internal cognition is now within reach. Embedding such interpretability tools within LLM infrastructures is crucial to achieving proactive risk mitigation rather than reactive crisis management.

Beyond extracting and interpreting fixed circuits, another frontier involves enabling models to dynamically assemble and route their own explanation strategies.

\paragraph{\textbf{Interpretability notes on Amodei's argumentation}}\cite{amodei2025interpretability}.
Although this reflection outlines the conceptual foundation for audience-sensitive XAI, several technical and methodological limitations remain that must be addressed in future research and implementation efforts:

\begin{itemize}
    \item \textbf{Opacity and Alignment Risk:} Amodei stresses that many concerns about alignment, such as deception, power seeking, and emerging behaviors, stem from our inability to observe internal mechanisms. Without interpretability, these risks remain undetectable or unverifiable.
    
    \item \textbf{Dynamic Explanation Mechanisms:} The frontier of interpretability involves enabling models not only to be analyzed post hoc but to autonomously construct and route explanatory strategies, opening possibilities for self-monitoring systems.

    \item \textbf{Regulatory and Scientific Barriers:} Amodei highlights how the lack of structured, auditable output prevents LLMs from being adopted in legally restricted domains, such as loan approval or scientific research, where an explanation is a requirement.

    \item \textbf{Geopolitical and Deployment Urgency:} Describing a near-future scenario where models may exhibit national-scale cognitive capabilities (“a country of geniuses in a datacenter”), Amodei warns that interpretability must mature before such power is deployed, especially to ensure safe use by democratic nations rather than autocracies.
\end{itemize}

\section{Designing Actionable and Governable XAI: Challenges and Research Frontiers}
\label{sec:challenges}

As LLMs increasingly serve as decision support tools in socially consequential domains, the design of their explainability systems must evolve beyond technical metrics to address multidimensional demands. This section outlines the pressing \textit{needs and opportunities} for explainable LLMs, including task alignment, user interaction, and transparency infrastructure, followed by a forward-looking perspective on what it means to develop \textit{governance-ready} XAI. We argue that explainability must be structurally embedded-supporting oversight, contestability, and institutional trustworthiness as foundation models scale toward agentic and autonomous capabilities.

\subsection{Needs and Opportunities of XAI for LLMs}
\label{sec:Needs}

As LLM deployment expands into domains where decisions affect rights, safety, and well-being, explainability is no longer optional, it is a condition for institutional legitimacy and public trust. This section reflects on current challenges, emerging methodologies, and the path forward to designing XAI that is epistemically robust and socially aligned.

\paragraph{1. From performance to functional intelligence.}
The success of LLM in producing fluent outputs has led to an overemphasis on performance metrics, such as accuracy or perplexity. However, in real-world domains, intelligibility-free correctness undermines user agency. We argue that the goal of XAI is not total transparency, but \textit{functional intelligibility}—the ability for different stakeholders to ask appropriate questions and receive context-sensitive actionable responses.

\paragraph{2. Challenges of scaling and opaqueness.}
Architectural complexity of LLMs introduces substantial barriers to interpretability. Their emergent behavior, vast parameter space, and prompt sensitivity make it difficult to explain or audit decisions. These issues are compounded by the absence of standardized evaluation metrics and explanation formats and by the need to produce layered explanations tailored to distinct audiences such as patients, regulators, or developers.

\paragraph{3. Technical and methodological gaps.} 
While this reflection outlines the conceptual foundation for audience-sensitive XAI, several technical and methodological limitations remain that must be addressed in future research and implementation efforts:
\begin{itemize}
    \item \textit{Lack of empirical grounding}: This reflection is conceptual and normative, lacking implementation-level evaluations.
    \item \textit{Absent modeling of stakeholder profiles}: Explanation types are not yet systematically mapped to audience needs.
    \item \textit{Evaluation inconsistency}: Metrics such as faithfulness, plausibility, and robustness remain unevenly applied in practice.
    \item \textit{Underdeveloped domain-specific tailoring}: Most XAI tools assume general-purpose LLMs rather than specialized applications in law, medicine, or education. 
\end{itemize}

\paragraph{4. Adaptive tool-use and self-modulating explanation systems.}
A promising path for future explainability involves not only making LLMs interpretable, but also equipping them to generate and adapt their own explanation pipelines based on situational needs. Toolformer~\cite{schick2023toolformer} exemplifies this shift: a language model that learns self-supervised to decide when, how and why to use external tools such as calculators, search engines, and question-response systems. This architecture enables a model to act as an explanation-enhanced agent, choosing its own tool invocation strategy based on internal uncertainty or performance signals. Although Toolformer was designed for factual enhancement and task solving, the underlying mechanism suggests a future in which LLMs autonomously determine which explanation modules to invoke depending on the user’s role, query complexity, or regulatory demands.  

\paragraph{5. Toward Personalized and Adaptive Explanations.} 
Future XAI systems must dynamically tailor explanations to individual users based on their expertise, goals, cognitive preferences, and real-time feedback. Static, one-size-fits-all explanations are insufficient in high-risk domains like medicine, law, and finance, where end-users range from domain experts and auditors to laypeople and affected individuals. Personalized XAI involves adjusting not only \emph{depth} (e.g., surface versus mechanistic reasoning), but also \emph{style} (e.g., narrative versus analytic) and \emph{modality} (e.g., text, visualization, interaction) of the explanation.

Achieving this adaptivity requires integrated user modeling, tracking user knowledge states, decision history, and explanation preferences, as well as interaction-based learning, where the system refines its explanation strategy through dialogic engagement. Techniques such as explanation intent classification, explanation refinement prompts, and interactive counterfactuals enable the co-construction of meaning between user and model. Dialogic XAI interfaces, powered by LLMs, can allow users to interrogate and negotiate terms of understanding, creating a feedback loop between explanation generation and comprehension.

In addition, such personalization is essential to promote contestability, trust calibration, and effective oversight. For example, a regulator may require structured evidentiary traces for a legal audit, while a patient may need empathetic narrative forms to support informed consent. Adaptive XAI also opens opportunities for accessibility: adapting explanations to neurodiverse users or multilingual populations through modality switching and simplification strategies.

\paragraph{6. Small open-weight language models and low-resource governance.}
As Small Language Models (SLMs)~\cite{lu2024small,subramanian2025small,wang2024comprehensive} gain traction in edge computing, embedded AI, and resource-constrained environments, their importance as deployable, interpretable, and privacy-preserving alternatives to large-scale LLMs is becoming increasingly clear. Although foundation models such as GPT-4 or Claude require significant computational infrastructure and centralized control, SLMs enable greater autonomy, energy efficiency, and deployment on devices with limited connectivity or hardware. Their reduced capacity improves explainability by lowering architectural complexity, thus improving transparency and auditability.

Wang et al.~\cite{wang2024comprehensive} provide a comprehensive taxonomy of the trustworthiness of SLM, emphasizing five core dimensions: robustness, privacy, reliability, safety, and fairness. Their findings suggest that SLMs can sometimes outperform larger LLMs in fairness and privacy settings, while also being more amenable to governance in sensitive domains. However, they also highlight that many state-of-the-art XAI methods, optimized for large models, struggle to generalize to SLM architectures, revealing a gap in trustworthiness evaluation and explanation for these models.

Future research must therefore prioritize lightweight, cognitively aligned XAI techniques tailored for SLMs, especially those operating in under-resourced regions or high-stakes environments. In addition, SLMs can enable local communities to maintain control over domain-specific AI deployments without relying on centralized infrastructures. This aligns with the larger goals of technological sovereignty, data minimization, and decentralized AI governance.

However, the simplicity of SLMs introduces new explainability risks: oversimplified logic, reduced context modeling, and difficulty in capturing nuanced dependencies. Therefore, the challenge lies in designing XAI frameworks that balance interpretability with performance in constrained settings, while also satisfying the emerging trust benchmarks outlined in~\cite{wang2024comprehensive}.

Subramanian et al.~\cite{subramanian2025small} further reinforce this trajectory by showcasing how SLMs can rival or even outperform larger LLMs across a range of tasks when paired with high-quality training techniques, optimized architectures, and efficient deployment strategies. Their survey highlights that task-agnostic and task-specific SLMs, spanning 1B to 8B parameters, can be designed to retain reasoning capabilities while significantly reducing computational burden. Notably, they emphasize that the “effective size” of SLMs—measured in performance equivalence to much larger models—depends heavily on factors such as data curation, modular training, and explanation tuning. This underlines a key XAI implication: smaller models, when deliberately trained and architected, offer not only interpretability advantages but also opportunities for audience-sensitive explanation. As such, SLMs serve as fertile ground for developing explainability methods that are cognitively aligned, efficient, and capable of deployment in decentralized governance frameworks—bridging the performance gap while enhancing user trust and oversight.

\paragraph{7. Emerging methodologies towards actionable frameworks.} Recent research opens promising directions to address these challenges.\\

\textbf{Implicit bias mechanisms in self-attention: The role of massive activations.}
Recent work by Sun et al. \cite{sun2024massive} identifies a surprising and critical phenomenon in the internal representations of LLMs: a very small number of activations reach values up to 10,000 times greater than the median, remaining largely constant across inputs. These so-called \textit{massive activations} function as learned bias terms that concentrate attention on specific tokens, usually the start of a sequence or structural delimiters, and serve as implicit scaffolds for the computation of self-attention. Their existence and effect challenge prevailing assumptions about input-driven reasoning and suggest that LLMs encode static architectural anchors not exposed by typical explanation tools. This deepens the need for mechanistic transparency, especially within RED XAI frameworks. Moreover, Sun et al. show that augmenting LLMs with explicit attention biases can eliminate these massive activations without degrading performance, offering a compelling direction for explainable-by-design architectures. These insights expand our call for XAI systems that uncover and address internal bias structures to ensure both epistemic fidelity and governance readiness.\\

\textbf{Causality and Trade-Offs.} Binkyte et al.~\cite{binkyte2025causality} argue that causality offers a principled lens for navigating the trade-offs between key goals in trustworthy machine learning, including fairness, privacy, robustness, accuracy, and explainability. Rather than optimizing these objectives in isolation, causal reasoning enables a structured analysis of their intersections, revealing both tensions and synergies. For example, causal models can distinguish between discriminatory and predictive features, facilitating interventions that preserve fairness without sacrificing accuracy. They can also be used to implement privacy-preserving techniques that minimize accuracy loss by disrupting only privacy-relevant pathways. This systemic perspective offers a powerful foundation for reconciling contestable AI decisions and ensuring that sensitive attributes do not exert undue influence over outcomes.

Causal frameworks such as structural Causal Models (SCMs) provide formal tools for modeling these trade-offs. SCMs allow interventions via the do-calculus to isolate direct and indirect causal effects, enabling precise counterfactual analyses. This is especially useful in high-risk domains such as healthcare or finance, where understanding \textit{why} behind a model's prediction is as important as the prediction itself. Moreover, explainability is deeply embedded in the causal view: counterfactual explanations and path-specific effects help auditors and stakeholders trace the logic behind decisions, supporting legal mandates such as GDPR's "right to explanation."

Liu et al.~\cite{liu2403large} expand this perspective to the realm of large language models (LLMs), showing that causality is not only instrumental in improving LLM trustworthiness, but also benefits from LLMs' generative capabilities. Their survey maps two-way synergies: (i) causal methods can reduce hallucination, enhance robustness, and improve the fairness of LLM predictions; (ii) LLMs can be leveraged to automate causal inference tasks such as treatment effect estimation, counterfactual generation, and causal graph discovery. For example, LLMs have been used to identify causal terms in unstructured text, augment small datasets with plausible counterfactuals, and support domain expert reasoning through causal querying.

Together, these studies highlight a critical opportunity for XAI: embedding causal reasoning in both the architecture of the foundation models and their interfaces with users. This includes using causal representations for fine-tuning, enforcing fairness through counterfactual regularization, and deploying causal audits to trace how particular training examples affect output. These methods move explanation from surface-level salience to deep causal attribution, making XAI not just descriptive, but actionable.

Recent research interprets the self-attention mechanism in Transformers as the estimation of a structural equation model (SEM) \cite{rohekar2023causal}, which allows the discovery of causal structures over input tokens without additional training. This approach enables zero-shot causal discovery, providing more robust and context-aware explanations

\textbf{Conversational Complexity as Risk Signal.} Burden et al. \cite{burden2024conversational} propose algorithmic measures such as \emph{Conversational Length (CL)} and \emph{Conversational Complexity (CC)} to quantify how easily a user can elicit harmful behavior from an LLM. Rather than treating risk as a static property of individual prompts or outputs, their framework frames it as a dynamic and emergent property of user-model interactions over time. This approach shifts the focus of XAI from explaining isolated outputs to characterizing the structure and effort required to navigate toward unsafe outcomes.

Conversational complexity is grounded in Kolmogorov complexity \cite{li2008introduction} and reflects the minimal informational effort required to steer a model toward a harmful response. A conversation with low CC and CL suggests that the model is vulnerable to trivial or easily accessible attacks. Conversely, a high CC may indicate that more elaborate strategies or obfuscations are necessary. In this way, CC and CL become explanatory tools for auditing \emph{how} and \emph{how easily} an LLM can be exploited, complementing internal interpretability techniques such as feature attribution or circuit tracing.

This perspective introduces several novel opportunities for XAI research and deployment:
\begin{itemize}
    \item \emph{Effort-Aware Safety Metrics:} Integrating CC and CL into red-teaming and safety evaluations allows a prioritization of conversations that pose the greatest real-world risk, those requiring minimal user effort.
    \item \emph{Interaction-Centric Explanations:} XAI systems can evolve from static saliency maps to dialogue-aware explanations that describe risky conversational patterns or escalation paths.
    \item \emph{Real-Time Risk Forecasting:} Monitoring complexity metrics in deployed systems can enable anticipatory interventions, identifying conversational trajectories before harmful content is generated.
    \item \emph{Tooling for Auditors and Policymakers:} These metrics provide structured and quantifiable evidence for audit logs, regulatory reporting, or compliance with frameworks such as the EU AI Act.
\end{itemize}

By embedding these interaction-level signals within broader interpretability pipelines, explainability moves closer to fulfilling its societal mandate, not only explaining what the model did, but revealing how and why it could be led astray. The framework of Burden et al. thus contributes a critical missing piece to operationalizing XAI in real-world safety workflows.

\textbf{Explanation-Answer Discrepancy.} Taghanaki and Monteiro~\cite{taghanaki2025explain} introduce the Explain-Query-Test (EQT) framework to assess whether LLMs can reason consistently across the full arc of explanation and question-answering. The EQT method decouples explanation generation from subsequent question answering by prompting a model to (i) explain a concept, (ii) generate multiple-choice questions from its own explanation, and (iii) answer those questions without referencing the original explanation. Their results expose a systematic discrepancy: LLMs that produce fluent and seemingly coherent explanations often fail to correctly answer questions based on those very explanations.

This misalignment suggests that the plausibility of the explanation does not equate to genuine comprehension or the coherence of internal knowledge. The authors formalize this issue through a self-contained feedback loop, in which explanation, question generation, and answering are treated as successive transformations of internal model knowledge. They propose \textit{Accloop}, a full-loop accuracy metric that reflects how faithfully the answers of a model align with its own explanations. A high \textit{Accloop} score implies internal consistency and integrity of reasoning, while a low score signals representational or inferential gaps within the model.

EQT thus offers a novel XAI lens, shifting attention from explanation quality in isolation to explanation consistency as a proxy for model understanding. The framework also introduces the \textit{Answer Consistency Score (ACS)}, which measures the answer stability in the asked questions, further highlighting that even minor variations in phrasing can cause models to break the logical coherence. These insights open important opportunities for XAI research: namely, to develop explanation methods that are not only plausible and faithful, but also \emph{verifiably consistent} in reasoning tasks. EQT provides a practical and testable scaffold for aligning model explanations with their downstream behavior, which is particularly critical in high-risk applications that require both interpretability and reliability.

\paragraph{Summary and Research Imperatives.}
This reflection should be seen as a conceptual framework and a call to action for empirically grounded, audience-aware, and interdisciplinary XAI research. Future work must:
\begin{itemize}
    \item Develop and benchmark adaptive explanation systems.
    \item Align the explanation design with the goals and regulatory requirements of the stakeholders.
    \item Translate technical innovation into societal and ethical intelligibility.
\end{itemize}

The path forward demands not only better models, but better socio-technical alignment, so that explainability becomes a mechanism of trust, not just a trace of function.

\subsection{Toward a Governance-Ready Future for XAI in LLMs}
\label{sec:Research}

Explainability must function not only as a tool for user understanding, but also as a critical interface between AI systems and institutional governance. In high-risk, regulated domains—such as healthcare, finance, and the public sector—explanations must satisfy formal requirements: legal accountability (e.g., GDPR Article 22), compliance auditing, and transparent decision documentation. Future research should expand the role of explanations beyond individual intelligibility to include their function as legal instruments, audit-ready artifacts, and participatory feedback mechanisms. This governance-ready perspective reframes XAI as part of an institutional control layer, one that enables contestability, supports regulatory oversight, and reinforces the procedural legitimacy of AI-assisted decision-making.

Building on the challenges and opportunities outlined in this reflection, I identify several strategic directions for future research and governance in explainable LLMs. These frontiers call for interdisciplinary collaboration in technical, legal, ethical, and institutional domains. I pay attention to opacity as a governance challenge (1), infrastructure and standards (2–4), governance interfaces (5-6), interaction and adaptation (7-9), and scale and context (10-11). 

\paragraph{1. Opacity as a governance challenge.}
LLMs exhibit structural opacity due to their complex architectures and emergent behaviors. As emphasized throughout this reflection, opacity is not inherently unethical: It becomes problematic only when left unmanaged. Future research must treat opacity not as a defect to be eliminated, but as a design condition to be ethically scaffolded through layered, role-sensitive explainability and institutional oversight \cite{fraser2022ai,herrera2025reflections}. This rethinking implies that XAI should serve as a governance interface capable of rendering opaque systems intelligible, contestable, and procedurally accountable. Efforts must prioritize explanation frameworks that shift trust from visibility to governability, especially in high-risk and legally regulated domains.

\paragraph{2. Theory-to-practice bridges.}
Future work must translate normative XAI principles, such as contestability, stakeholder calibration, and functional intelligibility, into deployable systems. This includes building explanation modules that are formally aligned with the roles of the audience and legal expectations.

\paragraph{3. Evaluation and measurement ecosystems}
While progress has been made in defining metrics (e.g., faithfulness, robustness, contrastivity), the field still lacks shared standards and realistic benchmarks for multirole, multidomain explanation quality. Community-led efforts must build evaluation suites that simulate real-world use cases in medicine, law, education, and finance.

\paragraph{4. Auditable and modular architectures.}
Inspired by the mechanistic interpretability, XAI research should move toward the construction of modular subnets or explanation circuits that can be independently audited, updated, or governed. This includes explainable-by-design architectures for both general-purpose and domain-specialized LLMs.

\paragraph{5. Explanation as a governance interface.}
Explainability must extend beyond user-centric transparency to serve as a functional layer in institutional governance. As AI systems are increasingly deployed in high-stakes public and regulated domains, explanations should operate as legal artifacts (e.g., for Article 22 GDPR compliance), regulatory filings, and formal audit records. Future research should explore how explanation systems can support procedural fairness, enable contestability, and provide participatory feedback mechanisms, thus transforming XAI into a bridge between algorithmic logic and institutional accountability.

\paragraph{6. Causal AI for governance-ready explainability.}
Future research should deepen the integration of causal reasoning frameworks into XAI for LLMs. Unlike traditional correlational explanations, causal AI can identify cause-effect relationships in model outputs, enabling both technical introspection and user-facing counterfactuals. Research priorities include developing scalable causal discovery methods within LLM architectures (e.g. self-attention as Structural Equation Models), building user interfaces for causal counterfactual exploration, and establishing causal audit trails for regulatory compliance. Governance frameworks must recognize causal explanations as vital instruments for ensuring accountability, transparency, and human-centered decision justification in critical domains such as healthcare, education, and law.

\paragraph{6. Personalized, adaptive and dialogic XAI,}
Explanations must evolve into adaptive, interaction-aware systems that account for user profiles, decision context, and feedback loops. Research should advance dialogic interfaces, interactive tracing, and explanation personalization in dynamic real-time environments.

\paragraph{7. Toward self-assembling explanation pipelines for AI aligned with humans.}
Future XAI infrastructures could therefore include not only prescribed explanation formats, but also self-assembling pipelines where models strategically route between modules for narrative generation, fact checking, and justification, enabling hybrid intelligibility that evolves alongside user needs.

\paragraph{8. The role of LLMs in human-AI collaboration.} 
Looking ahead, building a governance-ready future for XAI requires reimagining LLMs not as static explainers, but as interactive collaborators within complex human decision-making ecosystems. Future research must explore how LLMs can dynamically adapt their communicative strategies based on user experience, task complexity, and cognitive context. Rather than functioning as fixed oracles, LLMs should evolve into responsive teammates capable of negotiating uncertainty, seeking clarification, and guiding users through domain-specific reasoning processes. This involves not only improved explanation generation, but also interactive scaffolding that enhances user comprehension and agency. On the governance side, the emergence of LLMs as cognitive collaborators raises new questions about responsibility attribution, oversight mechanisms, and institutional accountability. As these systems influence decision making in sensitive domains, it becomes essential to design policies and evaluation frameworks that ensure that explanations are not only technically sound but also socially meaningful and procedurally trustworthy, fostering contestability, shared reasoning, and informed consent toward cointelligence.

\paragraph{10. Small language models and low resource governance}
As SLMs gain traction in edge and resource-constrained settings, future research must develop lightweight XAI techniques on-device. These should support governance and trust in decentralized, domain-specific, or underresourced regions.

\paragraph{11. Cross-domain validation and longitudinal studies.}
Future work should include longitudinal cross-sector case studies that evaluate the real-world impact of XAI practices on user trust, decision quality, and regulatory compliance. This empirical grounding is crucial to ensure that the explanation is not just symbolic but also socially functional.

\paragraph{Closing Note.}
These frontiers are not isolated technical puzzles. They are sociotechnical design challenges that demand hybrid thinking across law, machine learning, human-LLM collaboration, ethics, and public policy. As LLMs grow in influence. Explainability must evolve into a civic infrastructure that serves not just users, but societies. Governance-ready XAI is not just an implementation goal; it is the foundation upon which accountable, contestable, and human-aligned AI systems will stand.

\section{Conclusion and Reflections}
\label{sec:Conclusion}

This paper has argued that explainability in LLMs is not a secondary concern but a foundational requirement for epistemic legitimacy, democratic accountability, and human-LLM alignment. As LLMs increasingly mediate decisions that affect people’s rights, safety, and dignity, explainability becomes a civic and institutional infrastructure, one that must bridge the gap between technical design and societal values.

I propose that an effective XAI system for LLMs must be:
\begin{itemize}
    \item \textit{Audience-aware:} Tailored to the cognitive needs, roles, and goals of its diverse stakeholders.
    \item \textit{Operationally embedded:} Integrated into development pipelines to support alignment, debugging, and oversight.
    \item \textit{Governance-ready:} Aligned with legal, ethical, and democratic principles.
\end{itemize}

Opacity in LLMs should not merely be treated as a flaw to be eliminated, but as a structural reality to be ethically scaffolded. This demands layered, role-sensitive explanations that support real-world use: from audit trails for regulators to metaphoric analogies for users and counterfactual reasoning for developers. Explainability, in this sense, is not just a functional requirement, but a democratic one.

As LLMs are increasingly integrated into collaborative settings, such as education, medicine, and decision support, their role is changing from passive information providers to active cognitive partners. This transformation aligns with the vision of \textit{“cointelligence”} \cite{mollick2024co}, where human-AI interaction is dialogic, trust-based, and mutually adaptive. In human-AI collaboration, LLMs must augment human decision-making rather than simply automate it. In these scenarios, explainability becomes not only a means of understanding, but also a mechanism for skill building, trust calibration, and shared reasoning.

It is important to clarify that this paper does not aim to provide an exhaustive review of all existing XAI techniques for LLMs. Rather, it offers a reflective synthesis that builds on and complements several recent surveys and overviews of LLM explainability. The focus is not on enumerating methods, but on distilling key conceptual tensions, highlighting overlooked dimensions, and advancing a human-centered perspective grounded in the core dimensions of \textit{faithfulness}, \textit{truthfulness}, \textit{plausibility}, and \textit{contrastivity}.

Several recent surveys inform and contextualize this reflection. Luo and Specia~\cite{luo2024understanding} present an evolving taxonomy of XAI techniques for LLMs, highlighting the interaction between interpretability goals and model complexity. Zhao et al.~\cite{zhao2024explainability} provide a comprehensive review of the system level, identifying open challenges in model transparency, evaluation protocols, and stakeholder interaction. Mumuni and Mumuni~\cite{mumuni2025explainable} trace the evolution from inherently interpretable models to the post-hoc explainability paradigms required for LLMs. Bilal et al.~\cite{bilal2025llms} explore the dual role of LLMs as both explainees and explainers, highlighting the increasing relevance of LLM-generated explanations in multi-agent and tool-augmented settings. Cambria et al.~\cite{cambria2024xai} offer a broader lens on the convergence of XAI and LLM development, calling for a more integrated research agenda that unites natural language understanding with user-aligned transparency.

Furthermore, Ji et al.~\cite{ji2023survey} survey hallucination phenomena in natural language generation, emphasizing the importance of grounding outputs in verifiable facts, an issue central to our treatment of \textit{truthfulness}. Lin et al.~\cite{lin2024towards} review strategies for reducing bias and hallucinations in LLMs, providing valuable grounding for governance-ready XAI. Liu et al.~\cite{liu2403large} examine how causal inference techniques can enhance the interpretability and fairness of LLM outputs, supporting our case for explanation strategies that go beyond correlation. Xu et al.~\cite{xu2024contam} map the underexplored risks of benchmark data contamination in LLM training pipelines, pointing to the need for explainability mechanisms that account for provenance and training-set attribution. Finally, Rai et al.~\cite{rai2025practical} focus on mechanistic interpretability techniques such as circuit tracing and feature attribution, offering a workflow-oriented roadmap that complements our emphasis on layered, audience-aware XAI.

Together, these works highlight the growing multidimensionality of explainability in the LLM era and reinforce the need for explanation frameworks that are not only technically sound, but also socially intelligible and stakeholder-aligned. This paper addresses this need by organizing the landscape through four core dimensions—\textit{faithfulness}, \textit{truthfulness}, \textit{plausibility}, and \textit{contrastivity}—to guide the development of adaptive, evaluable, and trustworthy XAI for LLM.

\paragraph{Final Reflection.}  
This reflection was motivated not by abstract technical curiosity, but by a deep human imperative. How can we ensure that the intelligence we build remains in the service of those it is meant to empower? The metaphor of "making sense of the unsensible" captures a moral and epistemic challenge at the core of the LLM era. These systems increasingly shape the world, but do they help us understand it?

A central contribution of this reflection is the articulation of four core dimensions essential to explainability in LLMs: \textit{faithfulness}, \textit{truthfulness}, \textit{plausibility}, and \textit{contrastivity}. These dimensions are not only conceptual touchstones but also design constraints that expose the inherent tensions in building socially responsible explanation systems. Although each dimension aims at a distinct evaluative lens: internal reasoning, external factuality, human interpretability, and comparative justification—they must be balanced to produce explanations that are technically grounded, cognitively accessible, and contextually meaningful. We argue that explainability must go beyond surface-level justifications to support epistemic clarity, regulatory compliance, and audience-specific intelligibility, especially in the high-risk domains where LLMs are increasingly deployed.

In this light, explainability is more than a design artifact. It is a social contract between AI developers and the communities in which their systems impact. It serves as the channel through which AI becomes contestable, traceable, and just, thus supporting human-LLM decision-making. Without it, we risk not only miscommunication but also institutional breakdowns in trust and justice, potentially leading to the uncritical automation of human judgment~\cite{alam2024automation}. As Alam et al.~\cite{alam2024automation} emphasize, the path forward in high-impact domains lies not in full automation but in human-centered augmentation, where NextGen AI systems enhance expert reasoning, accountability, and collaborative decision making, rather than replacing them.

The future of human-AI collaboration does not lie in models that simply perform well; it lies in models that can be understood, questioned, and governed. Let this work serve as a call to action: for researchers to build explainability into the core of AI architectures; for policymakers to demand intelligibility in systems that affect lives; and for society to refuse the normalization of lack of opacity governance in AI systems.

\textit{ Making sense of the unsensible} is not just understanding machines; it reaffirms our commitment to human dignity, autonomy, and shared responsibility in the age of intelligent systems.

This work is not a conclusion, but a call to action: for researchers, developers, and policymakers to rethink XAI for LLMs as a shared interdisciplinary undertaking rooted in the lived realities of human experience.

\begin{quote}
  \textit{In making sense of the unsensible, we do not merely clarify machines; we shape how they support and extend human decision making, ensuring that the intelligence we create remains aligned with human values and in service of dignity, agency, and shared progress.}

\end{quote}

\noindent Key takeaways.
\begin{itemize}
  \item Explainability must evolve beyond static output to interactive, context-sensitive systems.
  \item Trust, contestability, and human oversight are not by-products, but primary functions of modern XAI.
  \item Future XAI systems must be judged not only by technical fidelity but also by how they enable action, accountability, and user agency.
  \item An effective LLM explainability must balance four core dimensions—\textit{faithfulness}, \textit{truthfulness}, \textit{plausibility}, and \textit{contrastivity}—to ensure explanations are technically grounded, factually accurate, cognitively accessible, and contextually meaningful.
  \item In collaborative contexts, XAI should enable LLMs to enhance human decision-making capabilities and support more effectively alongside cointelligence reasoning.
\end{itemize}

\subsubsection*{Acknowledgements} 
The author thanks Luciano Sánchez, Jose Hernández-Orallo (Spain), and Dionis López (Cuba) for their reading and comments.

This research results from the Strategic Project IAFER-Cib (C074/23), as a result of the collaboration agreement signed between the National Institute of Cybersecurity (INCIBE) and the University of Granada. This initiative is carried out within the framework of the Recovery, Transformation and Resilience Plan funds, financed by the European Union (Next Generation).

\section*{Declaration of AI-assisted technologies in the writing process}

During the preparation of this work, the author used large-language models to improve the readability and language of the manuscript. After using this tool/service, the author reviewed and edited the content as needed and assumed full responsibility for the content of the published article.

\bibliographystyle{apacite}
\bibliography{xai_llm_refs2}

\begin{thebibliography}{}

\bibitem [\protect \citeauthoryear {%
Alam%
\ \protect \BOthers {.}}{%
Alam%
\ \protect \BOthers {.}}{%
{\protect \APACyear {2024}}%
}]{%
alam2024automation}
\APACinsertmetastar {%
alam2024automation}%
\begin{APACrefauthors}%
Alam, F.%
, Lentsch, A.%
, Yu, N.%
, Barmack, S.%
, Kim, S.%
, Acemoglu, D.%
\BDBL {}Ahmed, F.%
\end{APACrefauthors}%
\unskip\
\newblock
\APACrefYearMonthDay{2024}{}{}.
\newblock
\APACrefbtitle {From Automation to Augmentation: Redefining Engineering Design and Manufacturing in the Age of NextGen-{AI}} {From automation to augmentation: Redefining engineering design and manufacturing in the age of nextgen-{AI}}\ \APACbVolEdTR{}{\BTR{}}.
\newblock
\APACaddressInstitution{}{MIT}.
\PrintBackRefs{\CurrentBib}

\bibitem [\protect \citeauthoryear {%
Ameisen%
\ \protect \BOthers {.}}{%
Ameisen%
\ \protect \BOthers {.}}{%
{\protect \APACyear {2025}}%
}]{%
ameisen2025circuit}
\APACinsertmetastar {%
ameisen2025circuit}%
\begin{APACrefauthors}%
Ameisen, E.%
, Lindsey, J.%
, Pearce, A.%
, Gurnee, W.%
, Turner, N\BPBI L.%
, Chen, B.%
\BDBL {}Batson, J.%
\end{APACrefauthors}%
\unskip\
\newblock
\APACrefYearMonthDay{2025}{}{}.
\newblock
\APACrefbtitle {Circuit Tracing: Revealing Computational Graphs in Language Models.} {Circuit tracing: Revealing computational graphs in language models.}
\newblock
\begin{APACrefURL} \url{https://transformer-circuits.pub/2025/attribution-graphs/methods.html} \end{APACrefURL}
\newblock
\APACrefnote{Anthropic Blog}
\PrintBackRefs{\CurrentBib}

\bibitem [\protect \citeauthoryear {%
Amodei%
}{%
Amodei%
}{%
{\protect \APACyear {2025}}%
}]{%
amodei2025interpretability}
\APACinsertmetastar {%
amodei2025interpretability}%
\begin{APACrefauthors}%
Amodei, D.%
\end{APACrefauthors}%
\unskip\
\newblock
\APACrefYearMonthDay{2025}{}{}.
\newblock
\APACrefbtitle {The Urgency of Interpretability.} {The urgency of interpretability.}
\newblock
\begin{APACrefURL} \url{https://www.darioamodei.com/post/the-urgency-of-interpretability#the-utility-of-interpretability} \end{APACrefURL}
\newblock
\APACrefnote{Anthropic}
\PrintBackRefs{\CurrentBib}

\bibitem [\protect \citeauthoryear {%
Arrieta%
\ \protect \BOthers {.}}{%
Arrieta%
\ \protect \BOthers {.}}{%
{\protect \APACyear {2020}}%
}]{%
arrieta2020explainable}
\APACinsertmetastar {%
arrieta2020explainable}%
\begin{APACrefauthors}%
Arrieta, A\BPBI B.%
, D{\'\i}az-Rodr{\'\i}guez, N.%
, Del~Ser, J.%
, Bennetot, A.%
, Tabik, S.%
, Barbado, A.%
\BDBL {}others%
\end{APACrefauthors}%
\unskip\
\newblock
\APACrefYearMonthDay{2020}{}{}.
\newblock
{\BBOQ}\APACrefatitle {Explainable Artificial Intelligence (XAI): Concepts, taxonomies, opportunities and challenges toward responsible AI} {Explainable artificial intelligence (xai): Concepts, taxonomies, opportunities and challenges toward responsible ai}.{\BBCQ}
\newblock
\APACjournalVolNumPages{Information Fusion}{58}{}{82--115}.
\PrintBackRefs{\CurrentBib}

\bibitem [\protect \citeauthoryear {%
Bertrand%
, Viard%
, Belloum%
, Eagan%
\BCBL {}\ \BBA {} Maxwell%
}{%
Bertrand%
\ \protect \BOthers {.}}{%
{\protect \APACyear {2023}}%
}]{%
bertrand2023selective}
\APACinsertmetastar {%
bertrand2023selective}%
\begin{APACrefauthors}%
Bertrand, A.%
, Viard, T.%
, Belloum, R.%
, Eagan, J\BPBI R.%
\BCBL {}\ \BBA {} Maxwell, W.%
\end{APACrefauthors}%
\unskip\
\newblock
\APACrefYearMonthDay{2023}{}{}.
\newblock
{\BBOQ}\APACrefatitle {On selective, mutable and dialogic XAI: A review of what users say about different types of interactive explanations} {On selective, mutable and dialogic xai: A review of what users say about different types of interactive explanations}.{\BBCQ}
\newblock
\BIn{} \APACrefbtitle {Proceedings of the 2023 CHI Conference on Human Factors in Computing Systems} {Proceedings of the 2023 chi conference on human factors in computing systems}\ (\BPGS\ 1--21).
\PrintBackRefs{\CurrentBib}

\bibitem [\protect \citeauthoryear {%
Biecek%
\ \BBA {} Samek%
}{%
Biecek%
\ \BBA {} Samek%
}{%
{\protect \APACyear {2024}}%
}]{%
biecek2024explain}
\APACinsertmetastar {%
biecek2024explain}%
\begin{APACrefauthors}%
Biecek, P.%
\BCBT {}\ \BBA {} Samek, W.%
\end{APACrefauthors}%
\unskip\
\newblock
\APACrefYearMonthDay{2024}{}{}.
\newblock
{\BBOQ}\APACrefatitle {Explain to Question not to Justify} {Explain to question not to justify}.{\BBCQ}
\newblock
\APACjournalVolNumPages{arXiv preprint arXiv:2402.13914}{}{}{}.
\PrintBackRefs{\CurrentBib}

\bibitem [\protect \citeauthoryear {%
Bilal%
, Ebert%
\BCBL {}\ \BBA {} Lin%
}{%
Bilal%
\ \protect \BOthers {.}}{%
{\protect \APACyear {2025}}%
}]{%
bilal2025llms}
\APACinsertmetastar {%
bilal2025llms}%
\begin{APACrefauthors}%
Bilal, A.%
, Ebert, D.%
\BCBL {}\ \BBA {} Lin, B.%
\end{APACrefauthors}%
\unskip\
\newblock
\APACrefYearMonthDay{2025}{}{}.
\newblock
{\BBOQ}\APACrefatitle {LLMs for Explainable AI: A Comprehensive Survey} {Llms for explainable ai: A comprehensive survey}.{\BBCQ}
\newblock
\APACjournalVolNumPages{arXiv preprint arXiv:2504.00125}{}{}{}.
\PrintBackRefs{\CurrentBib}

\bibitem [\protect \citeauthoryear {%
Binkyte%
\ \protect \BOthers {.}}{%
Binkyte%
\ \protect \BOthers {.}}{%
{\protect \APACyear {2025}}%
}]{%
binkyte2025causality}
\APACinsertmetastar {%
binkyte2025causality}%
\begin{APACrefauthors}%
Binkyte, R.%
, Sheth, I.%
, Jin, Z.%
, Havaei, M.%
, Sch{\"o}lkopf, B.%
\BCBL {}\ \BBA {} Fritz, M.%
\end{APACrefauthors}%
\unskip\
\newblock
\APACrefYearMonthDay{2025}{}{}.
\newblock
{\BBOQ}\APACrefatitle {Causality Is Key to Understand and Balance Multiple Goals in Trustworthy ML and Foundation Models} {Causality is key to understand and balance multiple goals in trustworthy ml and foundation models}.{\BBCQ}
\newblock
\APACjournalVolNumPages{arXiv preprint arXiv:2502.21123}{}{}{}.
\PrintBackRefs{\CurrentBib}

\bibitem [\protect \citeauthoryear {%
Blog%
}{%
Blog%
}{%
{\protect \APACyear {2025}}%
}]{%
Anthropic2025Tracing}
\APACinsertmetastar {%
Anthropic2025Tracing}%
\begin{APACrefauthors}%
Blog, A.%
\end{APACrefauthors}%
\unskip\
\newblock
\APACrefYearMonthDay{2025}{}{}.
\newblock
\APACrefbtitle {Tracing the thoughts of a large language model.} {Tracing the thoughts of a large language model.}
\newblock
\begin{APACrefURL} \url{https://www.anthropic.com/research/tracing-thoughts-language-model} \end{APACrefURL}
\newblock
\APACrefnote{Anthropic Blog}
\PrintBackRefs{\CurrentBib}

\bibitem [\protect \citeauthoryear {%
Boge%
}{%
Boge%
}{%
{\protect \APACyear {2022}}%
}]{%
boge2022two}
\APACinsertmetastar {%
boge2022two}%
\begin{APACrefauthors}%
Boge, F\BPBI J.%
\end{APACrefauthors}%
\unskip\
\newblock
\APACrefYearMonthDay{2022}{}{}.
\newblock
{\BBOQ}\APACrefatitle {Two dimensions of opacity and the deep learning predicament} {Two dimensions of opacity and the deep learning predicament}.{\BBCQ}
\newblock
\APACjournalVolNumPages{Minds and Machines}{32}{1}{43--75}.
\PrintBackRefs{\CurrentBib}

\bibitem [\protect \citeauthoryear {%
Burden%
, Cebrian%
\BCBL {}\ \BBA {} Hernandez-Orallo%
}{%
Burden%
\ \protect \BOthers {.}}{%
{\protect \APACyear {2024}}%
}]{%
burden2024conversational}
\APACinsertmetastar {%
burden2024conversational}%
\begin{APACrefauthors}%
Burden, J.%
, Cebrian, M.%
\BCBL {}\ \BBA {} Hernandez-Orallo, J.%
\end{APACrefauthors}%
\unskip\
\newblock
\APACrefYearMonthDay{2024}{}{}.
\newblock
{\BBOQ}\APACrefatitle {Conversational complexity for assessing risk in large language models} {Conversational complexity for assessing risk in large language models}.{\BBCQ}
\newblock
\APACjournalVolNumPages{arXiv preprint arXiv:2409.01247}{}{}{}.
\PrintBackRefs{\CurrentBib}

\bibitem [\protect \citeauthoryear {%
Cambria%
, Malandri%
, Mercorio%
, Nobani%
\BCBL {}\ \BBA {} Seveso%
}{%
Cambria%
\ \protect \BOthers {.}}{%
{\protect \APACyear {2024}}%
}]{%
cambria2024xai}
\APACinsertmetastar {%
cambria2024xai}%
\begin{APACrefauthors}%
Cambria, E.%
, Malandri, L.%
, Mercorio, F.%
, Nobani, N.%
\BCBL {}\ \BBA {} Seveso, A.%
\end{APACrefauthors}%
\unskip\
\newblock
\APACrefYearMonthDay{2024}{}{}.
\newblock
{\BBOQ}\APACrefatitle {Xai meets llms: A survey of the relation between explainable ai and large language models} {Xai meets llms: A survey of the relation between explainable ai and large language models}.{\BBCQ}
\newblock
\APACjournalVolNumPages{arXiv preprint arXiv:2407.15248}{}{}{}.
\PrintBackRefs{\CurrentBib}

\bibitem [\protect \citeauthoryear {%
Chen%
\ \protect \BOthers {.}}{%
Chen%
\ \protect \BOthers {.}}{%
{\protect \APACyear {2024}}%
}]{%
chen2024truth}
\APACinsertmetastar {%
chen2024truth}%
\begin{APACrefauthors}%
Chen, Z.%
, Sun, X.%
, Jiao, X.%
, Lian, F.%
, Kang, Z.%
, Wang, D.%
\BCBL {}\ \BBA {} Xu, C.%
\end{APACrefauthors}%
\unskip\
\newblock
\APACrefYearMonthDay{2024}{}{}.
\newblock
{\BBOQ}\APACrefatitle {Truth forest: Toward multi-scale truthfulness in large language models through intervention without tuning} {Truth forest: Toward multi-scale truthfulness in large language models through intervention without tuning}.{\BBCQ}
\newblock
\BIn{} \APACrefbtitle {Proceedings of the AAAI Conference on Artificial Intelligence} {Proceedings of the aaai conference on artificial intelligence}\ (\BVOL~38, \BPGS\ 20967--20974).
\PrintBackRefs{\CurrentBib}

\bibitem [\protect \citeauthoryear {%
de Carvalho~Souza%
, Souza%
\BCBL {}\ \BBA {} Weigang%
}{%
de Carvalho~Souza%
\ \protect \BOthers {.}}{%
{\protect \APACyear {2025}}%
}]{%
de2025unveiling}
\APACinsertmetastar {%
de2025unveiling}%
\begin{APACrefauthors}%
de Carvalho~Souza, M\BPBI E.%
, Souza, M\BPBI E\BPBI D\BPBI C.%
\BCBL {}\ \BBA {} Weigang, L.%
\end{APACrefauthors}%
\unskip\
\newblock
\APACrefYearMonthDay{2025}{}{}.
\newblock
{\BBOQ}\APACrefatitle {Unveiling the Black Box: The Significance of XAI in Making LLMs Transparent} {Unveiling the black box: The significance of xai in making llms transparent}.{\BBCQ}
\newblock
\APACjournalVolNumPages{Authorea Preprints}{}{}{}.
\PrintBackRefs{\CurrentBib}

\bibitem [\protect \citeauthoryear {%
D{\'\i}az-Rodr{\'\i}guez%
\ \protect \BOthers {.}}{%
D{\'\i}az-Rodr{\'\i}guez%
\ \protect \BOthers {.}}{%
{\protect \APACyear {2023}}%
}]{%
diaz2023connecting}
\APACinsertmetastar {%
diaz2023connecting}%
\begin{APACrefauthors}%
D{\'\i}az-Rodr{\'\i}guez, N.%
, Del~Ser, J.%
, Coeckelbergh, M.%
, L{\'o}pez~de Prado, M.%
, Herrera-Viedma, E.%
\BCBL {}\ \BBA {} Herrera, F.%
\end{APACrefauthors}%
\unskip\
\newblock
\APACrefYearMonthDay{2023}{}{}.
\newblock
{\BBOQ}\APACrefatitle {{Connecting the dots in trustworthy Artificial Intelligence: From {AI} principles, ethics, and key requirements to responsible {AI} systems and regulation}} {{Connecting the dots in trustworthy Artificial Intelligence: From {AI} principles, ethics, and key requirements to responsible {AI} systems and regulation}}.{\BBCQ}
\newblock
\APACjournalVolNumPages{Information Fusion}{99}{}{101896}.
\PrintBackRefs{\CurrentBib}

\bibitem [\protect \citeauthoryear {%
et al.%
}{%
et al.%
}{%
{\protect \APACyear {2025}}%
}]{%
Anthropic2025Biology}
\APACinsertmetastar {%
Anthropic2025Biology}%
\begin{APACrefauthors}%
et al., J\BPBI L.%
\end{APACrefauthors}%
\unskip\
\newblock
\APACrefYearMonthDay{2025}{}{}.
\newblock
\APACrefbtitle {On the Biology of a Large Language Model.} {On the biology of a large language model.}
\newblock
\begin{APACrefURL} \url{https://transformer-circuits.pub/2025/attribution-graphs/biology.html} \end{APACrefURL}
\newblock
\APACrefnote{Anthropic Paper}
\PrintBackRefs{\CurrentBib}

\bibitem [\protect \citeauthoryear {%
Fraser%
, Simcock%
\BCBL {}\ \BBA {} Snoswell%
}{%
Fraser%
\ \protect \BOthers {.}}{%
{\protect \APACyear {2022}}%
}]{%
fraser2022ai}
\APACinsertmetastar {%
fraser2022ai}%
\begin{APACrefauthors}%
Fraser, H.%
, Simcock, R.%
\BCBL {}\ \BBA {} Snoswell, A\BPBI J.%
\end{APACrefauthors}%
\unskip\
\newblock
\APACrefYearMonthDay{2022}{}{}.
\newblock
{\BBOQ}\APACrefatitle {Ai opacity and explainability in tort litigation} {Ai opacity and explainability in tort litigation}.{\BBCQ}
\newblock
\BIn{} \APACrefbtitle {Proceedings of the 2022 ACM Conference on Fairness, Accountability, and Transparency} {Proceedings of the 2022 acm conference on fairness, accountability, and transparency}\ (\BPGS\ 185--196).
\PrintBackRefs{\CurrentBib}

\bibitem [\protect \citeauthoryear {%
Garc{\'\i}a-Carrasco%
, Mat{\'e}%
\BCBL {}\ \BBA {} Trujillo%
}{%
Garc{\'\i}a-Carrasco%
\ \protect \BOthers {.}}{%
{\protect \APACyear {2024}}%
}]{%
garcia2024does}
\APACinsertmetastar {%
garcia2024does}%
\begin{APACrefauthors}%
Garc{\'\i}a-Carrasco, J.%
, Mat{\'e}, A.%
\BCBL {}\ \BBA {} Trujillo, J.%
\end{APACrefauthors}%
\unskip\
\newblock
\APACrefYearMonthDay{2024}{}{}.
\newblock
{\BBOQ}\APACrefatitle {How does gpt-2 predict acronyms? extracting and understanding a circuit via mechanistic interpretability} {How does gpt-2 predict acronyms? extracting and understanding a circuit via mechanistic interpretability}.{\BBCQ}
\newblock
\BIn{} \APACrefbtitle {International Conference on Artificial Intelligence and Statistics} {International conference on artificial intelligence and statistics}\ (\BPGS\ 3322--3330).
\PrintBackRefs{\CurrentBib}

\bibitem [\protect \citeauthoryear {%
Garc{\'\i}a-Carrasco%
, Mat{\'e}%
\BCBL {}\ \BBA {} Trujillo%
}{%
Garc{\'\i}a-Carrasco%
\ \protect \BOthers {.}}{%
{\protect \APACyear {2025}}%
}]{%
garcia2025extracting}
\APACinsertmetastar {%
garcia2025extracting}%
\begin{APACrefauthors}%
Garc{\'\i}a-Carrasco, J.%
, Mat{\'e}, A.%
\BCBL {}\ \BBA {} Trujillo, J.%
\end{APACrefauthors}%
\unskip\
\newblock
\APACrefYearMonthDay{2025}{}{}.
\newblock
{\BBOQ}\APACrefatitle {Extracting interpretable task-specific circuits from large language models for faster inference} {Extracting interpretable task-specific circuits from large language models for faster inference}.{\BBCQ}
\newblock
\BIn{} \APACrefbtitle {Proceedings of the AAAI Conference on Artificial Intelligence} {Proceedings of the aaai conference on artificial intelligence}\ (\BVOL~39, \BPGS\ 16772--16780).
\PrintBackRefs{\CurrentBib}

\bibitem [\protect \citeauthoryear {%
Haresamudram%
, Larsson%
\BCBL {}\ \BBA {} Heintz%
}{%
Haresamudram%
\ \protect \BOthers {.}}{%
{\protect \APACyear {2023}}%
}]{%
haresamudram2023three}
\APACinsertmetastar {%
haresamudram2023three}%
\begin{APACrefauthors}%
Haresamudram, K.%
, Larsson, S.%
\BCBL {}\ \BBA {} Heintz, F.%
\end{APACrefauthors}%
\unskip\
\newblock
\APACrefYearMonthDay{2023}{}{}.
\newblock
{\BBOQ}\APACrefatitle {{Three levels of {AI} transparency}} {{Three levels of {AI} transparency}}.{\BBCQ}
\newblock
\APACjournalVolNumPages{Computer}{56}{2}{93--100}.
\PrintBackRefs{\CurrentBib}

\bibitem [\protect \citeauthoryear {%
He%
, Aishwarya%
\BCBL {}\ \BBA {} Gadiraju%
}{%
He%
\ \protect \BOthers {.}}{%
{\protect \APACyear {2025}}%
}]{%
he2025conversational}
\APACinsertmetastar {%
he2025conversational}%
\begin{APACrefauthors}%
He, G.%
, Aishwarya, N.%
\BCBL {}\ \BBA {} Gadiraju, U.%
\end{APACrefauthors}%
\unskip\
\newblock
\APACrefYearMonthDay{2025}{}{}.
\newblock
{\BBOQ}\APACrefatitle {Is Conversational XAI All You Need? Human-AI Decision Making With a Conversational XAI Assistant} {Is conversational xai all you need? human-ai decision making with a conversational xai assistant}.{\BBCQ}
\newblock
\BIn{} \APACrefbtitle {Proceedings of the 30th International Conference on Intelligent User Interfaces} {Proceedings of the 30th international conference on intelligent user interfaces}\ (\BPGS\ 907--924).
\PrintBackRefs{\CurrentBib}

\bibitem [\protect \citeauthoryear {%
Herera%
\ \BBA {} Calderon%
}{%
Herera%
\ \BBA {} Calderon%
}{%
{\protect \APACyear {2025}}%
}]{%
Herrrera-Reyes2025}
\APACinsertmetastar {%
Herrrera-Reyes2025}%
\begin{APACrefauthors}%
Herera, F.%
\BCBT {}\ \BBA {} Calderon, R.%
\end{APACrefauthors}%
\unskip\
\newblock
\APACrefYearMonthDay{2025}{}{}.
\newblock
\APACrefbtitle {Opacity as a Feature, Not a Flaw: Trust, Governance Ethic, and Role-Sensitive Explainability in AI Systems.} {Opacity as a feature, not a flaw: Trust, governance ethic, and role-sensitive explainability in ai systems.}
\newblock
\APACrefnote{DaSCI Internal Report}
\PrintBackRefs{\CurrentBib}

\bibitem [\protect \citeauthoryear {%
Herrera%
}{%
Herrera%
}{%
{\protect \APACyear {2025}}%
}]{%
herrera2025reflections}
\APACinsertmetastar {%
herrera2025reflections}%
\begin{APACrefauthors}%
Herrera, F.%
\end{APACrefauthors}%
\unskip\
\newblock
\APACrefYearMonthDay{2025}{}{}.
\newblock
{\BBOQ}\APACrefatitle {Reflections and attentiveness on eXplainable Artificial Intelligence (XAI). The journey ahead from criticisms to human-AI collaboration} {Reflections and attentiveness on explainable artificial intelligence (xai). the journey ahead from criticisms to human-ai collaboration}.{\BBCQ}
\newblock
\APACjournalVolNumPages{Information Fusion}{}{}{103133}.
\PrintBackRefs{\CurrentBib}

\bibitem [\protect \citeauthoryear {%
Hsieh%
\ \protect \BOthers {.}}{%
Hsieh%
\ \protect \BOthers {.}}{%
{\protect \APACyear {2024}}%
}]{%
hsieh2024comprehensive}
\APACinsertmetastar {%
hsieh2024comprehensive}%
\begin{APACrefauthors}%
Hsieh, W.%
, Bi, Z.%
, Jiang, C.%
, Liu, J.%
, Peng, B.%
, Zhang, S.%
\BDBL {}others%
\end{APACrefauthors}%
\unskip\
\newblock
\APACrefYearMonthDay{2024}{}{}.
\newblock
{\BBOQ}\APACrefatitle {A comprehensive guide to explainable AI: from classical models to LLMs} {A comprehensive guide to explainable ai: from classical models to llms}.{\BBCQ}
\newblock
\APACjournalVolNumPages{arXiv preprint arXiv:2412.00800}{}{}{}.
\PrintBackRefs{\CurrentBib}

\bibitem [\protect \citeauthoryear {%
Jacovi%
\ \BBA {} Goldberg%
}{%
Jacovi%
\ \BBA {} Goldberg%
}{%
{\protect \APACyear {2020}}%
}]{%
jacovi2020towards}
\APACinsertmetastar {%
jacovi2020towards}%
\begin{APACrefauthors}%
Jacovi, A.%
\BCBT {}\ \BBA {} Goldberg, Y.%
\end{APACrefauthors}%
\unskip\
\newblock
\APACrefYearMonthDay{2020}{}{}.
\newblock
{\BBOQ}\APACrefatitle {Towards faithfully interpretable NLP systems: How should we define and evaluate faithfulness?} {Towards faithfully interpretable nlp systems: How should we define and evaluate faithfulness?}{\BBCQ}
\newblock
\APACjournalVolNumPages{arXiv preprint arXiv:2004.03685}{}{}{}.
\PrintBackRefs{\CurrentBib}

\bibitem [\protect \citeauthoryear {%
Ji%
\ \protect \BOthers {.}}{%
Ji%
\ \protect \BOthers {.}}{%
{\protect \APACyear {2023}}%
}]{%
ji2023survey}
\APACinsertmetastar {%
ji2023survey}%
\begin{APACrefauthors}%
Ji, Z.%
, Lee, N.%
, Frieske, R.%
, Yu, T.%
, Su, D.%
, Xu, Y.%
\BDBL {}Fung, P.%
\end{APACrefauthors}%
\unskip\
\newblock
\APACrefYearMonthDay{2023}{}{}.
\newblock
{\BBOQ}\APACrefatitle {Survey of hallucination in natural language generation} {Survey of hallucination in natural language generation}.{\BBCQ}
\newblock
\APACjournalVolNumPages{ACM computing surveys}{55}{12}{1--38}.
\PrintBackRefs{\CurrentBib}

\bibitem [\protect \citeauthoryear {%
Li%
, Vit{\'a}nyi%
\BCBL {}\ \protect \BOthers {.}}{%
Li%
\ \protect \BOthers {.}}{%
{\protect \APACyear {2008}}%
}]{%
li2008introduction}
\APACinsertmetastar {%
li2008introduction}%
\begin{APACrefauthors}%
Li, M.%
, Vit{\'a}nyi, P.%
\BCBL {}\ \BOthersPeriod {.}\end{APACrefauthors}%
\unskip\
\newblock
\APACrefYear{2008}.
\newblock
\APACrefbtitle {An introduction to Kolmogorov complexity and its applications} {An introduction to kolmogorov complexity and its applications}\ (\BVOL~3).
\newblock
\APACaddressPublisher{}{Springer}.
\PrintBackRefs{\CurrentBib}

\bibitem [\protect \citeauthoryear {%
S.~Lin%
, Hilton%
\BCBL {}\ \BBA {} Evans%
}{%
S.~Lin%
\ \protect \BOthers {.}}{%
{\protect \APACyear {2021}}%
}]{%
lin2021truthfulqa}
\APACinsertmetastar {%
lin2021truthfulqa}%
\begin{APACrefauthors}%
Lin, S.%
, Hilton, J.%
\BCBL {}\ \BBA {} Evans, O.%
\end{APACrefauthors}%
\unskip\
\newblock
\APACrefYearMonthDay{2021}{}{}.
\newblock
{\BBOQ}\APACrefatitle {Truthfulqa: Measuring how models mimic human falsehoods} {Truthfulqa: Measuring how models mimic human falsehoods}.{\BBCQ}
\newblock
\APACjournalVolNumPages{arXiv preprint arXiv:2109.07958}{}{}{}.
\PrintBackRefs{\CurrentBib}

\bibitem [\protect \citeauthoryear {%
Z.~Lin%
\ \protect \BOthers {.}}{%
Z.~Lin%
\ \protect \BOthers {.}}{%
{\protect \APACyear {2024}}%
}]{%
lin2024towards}
\APACinsertmetastar {%
lin2024towards}%
\begin{APACrefauthors}%
Lin, Z.%
, Guan, S.%
, Zhang, W.%
, Zhang, H.%
, Li, Y.%
\BCBL {}\ \BBA {} Zhang, H.%
\end{APACrefauthors}%
\unskip\
\newblock
\APACrefYearMonthDay{2024}{}{}.
\newblock
{\BBOQ}\APACrefatitle {Towards trustworthy LLMs: a review on debiasing and dehallucinating in large language models} {Towards trustworthy llms: a review on debiasing and dehallucinating in large language models}.{\BBCQ}
\newblock
\APACjournalVolNumPages{Artificial Intelligence Review}{57}{9}{243}.
\PrintBackRefs{\CurrentBib}

\bibitem [\protect \citeauthoryear {%
Liu%
\ \protect \BOthers {.}}{%
Liu%
\ \protect \BOthers {.}}{%
{\protect \APACyear {2024}}%
}]{%
liu2403large}
\APACinsertmetastar {%
liu2403large}%
\begin{APACrefauthors}%
Liu, X.%
, Xu, P.%
, Wu, J.%
, Yuan, J.%
, Yang, Y.%
, Zhou, Y.%
\BDBL {}others%
\end{APACrefauthors}%
\unskip\
\newblock
\APACrefYearMonthDay{2024}{}{}.
\newblock
{\BBOQ}\APACrefatitle {Large language models and causal inference in collaboration: A comprehensive survey. arXiv. 2024} {Large language models and causal inference in collaboration: A comprehensive survey. arxiv. 2024}.{\BBCQ}
\newblock
\APACjournalVolNumPages{arXiv preprint arXiv:2403.09606}{}{}{}.
\PrintBackRefs{\CurrentBib}

\bibitem [\protect \citeauthoryear {%
Longo%
\ \protect \BOthers {.}}{%
Longo%
\ \protect \BOthers {.}}{%
{\protect \APACyear {2024}}%
}]{%
longo2024explainable}
\APACinsertmetastar {%
longo2024explainable}%
\begin{APACrefauthors}%
Longo, L.%
, Brcic, M.%
, Cabitza, F.%
, Choi, J.%
, Confalonieri, R.%
, Del~Ser, J.%
\BDBL {}others%
\end{APACrefauthors}%
\unskip\
\newblock
\APACrefYearMonthDay{2024}{}{}.
\newblock
{\BBOQ}\APACrefatitle {Explainable Artificial Intelligence (XAI) 2.0: A manifesto of open challenges and interdisciplinary research directions} {Explainable artificial intelligence (xai) 2.0: A manifesto of open challenges and interdisciplinary research directions}.{\BBCQ}
\newblock
\APACjournalVolNumPages{Information Fusion}{106}{}{102301}.
\PrintBackRefs{\CurrentBib}

\bibitem [\protect \citeauthoryear {%
Lu%
\ \protect \BOthers {.}}{%
Lu%
\ \protect \BOthers {.}}{%
{\protect \APACyear {2024}}%
}]{%
lu2024small}
\APACinsertmetastar {%
lu2024small}%
\begin{APACrefauthors}%
Lu, Z.%
, Li, X.%
, Cai, D.%
, Yi, R.%
, Liu, F.%
, Zhang, X.%
\BDBL {}Xu, M.%
\end{APACrefauthors}%
\unskip\
\newblock
\APACrefYearMonthDay{2024}{}{}.
\newblock
{\BBOQ}\APACrefatitle {Small language models: Survey, measurements, and insights} {Small language models: Survey, measurements, and insights}.{\BBCQ}
\newblock
\APACjournalVolNumPages{arXiv preprint arXiv:2409.15790}{}{}{}.
\PrintBackRefs{\CurrentBib}

\bibitem [\protect \citeauthoryear {%
Luo%
\ \BBA {} Specia%
}{%
Luo%
\ \BBA {} Specia%
}{%
{\protect \APACyear {2024}}%
}]{%
luo2024understanding}
\APACinsertmetastar {%
luo2024understanding}%
\begin{APACrefauthors}%
Luo, H.%
\BCBT {}\ \BBA {} Specia, L.%
\end{APACrefauthors}%
\unskip\
\newblock
\APACrefYearMonthDay{2024}{}{}.
\newblock
{\BBOQ}\APACrefatitle {From understanding to utilization: A survey on explainability for large language models} {From understanding to utilization: A survey on explainability for large language models}.{\BBCQ}
\newblock
\APACjournalVolNumPages{arXiv preprint arXiv:2401.12874}{}{}{}.
\PrintBackRefs{\CurrentBib}

\bibitem [\protect \citeauthoryear {%
Mansurov%
, Sakip%
\BCBL {}\ \BBA {} Aji%
}{%
Mansurov%
\ \protect \BOthers {.}}{%
{\protect \APACyear {2024}}%
}]{%
mansurov2024data}
\APACinsertmetastar {%
mansurov2024data}%
\begin{APACrefauthors}%
Mansurov, J.%
, Sakip, A.%
\BCBL {}\ \BBA {} Aji, A\BPBI F.%
\end{APACrefauthors}%
\unskip\
\newblock
\APACrefYearMonthDay{2024}{}{}.
\newblock
{\BBOQ}\APACrefatitle {Data Laundering: Artificially Boosting Benchmark Results through Knowledge Distillation} {Data laundering: Artificially boosting benchmark results through knowledge distillation}.{\BBCQ}
\newblock
\APACjournalVolNumPages{arXiv preprint arXiv:2412.15255}{}{}{}.
\PrintBackRefs{\CurrentBib}

\bibitem [\protect \citeauthoryear {%
Mersha%
, Yigezu%
, Byun%
, Kalita%
\BCBL {}\ \protect \BOthers {.}}{%
Mersha%
\ \protect \BOthers {.}}{%
{\protect \APACyear {2025}}%
}]{%
mersha2025unified}
\APACinsertmetastar {%
mersha2025unified}%
\begin{APACrefauthors}%
Mersha, M\BPBI A.%
, Yigezu, M\BPBI G.%
, Byun, S.%
, Kalita, J.%
\BCBL {}\ \BOthersPeriod {.}\end{APACrefauthors}%
\unskip\
\newblock
\APACrefYearMonthDay{2025}{}{}.
\newblock
{\BBOQ}\APACrefatitle {A Unified Framework with Novel Metrics for Evaluating the Effectiveness of XAI Techniques in LLMs} {A unified framework with novel metrics for evaluating the effectiveness of xai techniques in llms}.{\BBCQ}
\newblock
\APACjournalVolNumPages{arXiv preprint arXiv:2503.05050}{}{}{}.
\PrintBackRefs{\CurrentBib}

\bibitem [\protect \citeauthoryear {%
Mollick%
}{%
Mollick%
}{%
{\protect \APACyear {2024}}%
}]{%
mollick2024co}
\APACinsertmetastar {%
mollick2024co}%
\begin{APACrefauthors}%
Mollick, E.%
\end{APACrefauthors}%
\unskip\
\newblock
\APACrefYear{2024}.
\newblock
\APACrefbtitle {Co-Intelligence: Living and Working with {AI}} {Co-intelligence: Living and working with {AI}}.
\newblock
\APACaddressPublisher{}{Random House UK}.
\PrintBackRefs{\CurrentBib}

\bibitem [\protect \citeauthoryear {%
Mumuni%
\ \BBA {} Mumuni%
}{%
Mumuni%
\ \BBA {} Mumuni%
}{%
{\protect \APACyear {2025}}%
}]{%
mumuni2025explainable}
\APACinsertmetastar {%
mumuni2025explainable}%
\begin{APACrefauthors}%
Mumuni, F.%
\BCBT {}\ \BBA {} Mumuni, A.%
\end{APACrefauthors}%
\unskip\
\newblock
\APACrefYearMonthDay{2025}{}{}.
\newblock
{\BBOQ}\APACrefatitle {Explainable artificial intelligence (XAI): from inherent explainability to large language models} {Explainable artificial intelligence (xai): from inherent explainability to large language models}.{\BBCQ}
\newblock
\APACjournalVolNumPages{arXiv preprint arXiv:2501.09967}{}{}{}.
\PrintBackRefs{\CurrentBib}

\bibitem [\protect \citeauthoryear {%
P{\l}onka%
, Kocot%
, Ho{\l}da%
, Daniec%
\BCBL {}\ \BBA {} Nawrat%
}{%
P{\l}onka%
\ \protect \BOthers {.}}{%
{\protect \APACyear {2025}}%
}]{%
plonka2025comparative}
\APACinsertmetastar {%
plonka2025comparative}%
\begin{APACrefauthors}%
P{\l}onka, M.%
, Kocot, K.%
, Ho{\l}da, K.%
, Daniec, K.%
\BCBL {}\ \BBA {} Nawrat, A.%
\end{APACrefauthors}%
\unskip\
\newblock
\APACrefYearMonthDay{2025}{}{}.
\newblock
{\BBOQ}\APACrefatitle {A comparative evaluation of the effectiveness of document splitters for large language models in legal contexts} {A comparative evaluation of the effectiveness of document splitters for large language models in legal contexts}.{\BBCQ}
\newblock
\APACjournalVolNumPages{Expert Systems with Applications}{}{}{126711}.
\PrintBackRefs{\CurrentBib}

\bibitem [\protect \citeauthoryear {%
Rai%
, Zhou%
, Feng%
, Saparov%
\BCBL {}\ \BBA {} Yao%
}{%
Rai%
\ \protect \BOthers {.}}{%
{\protect \APACyear {2025}}%
}]{%
rai2025practical}
\APACinsertmetastar {%
rai2025practical}%
\begin{APACrefauthors}%
Rai, D.%
, Zhou, Y.%
, Feng, S.%
, Saparov, A.%
\BCBL {}\ \BBA {} Yao, Z.%
\end{APACrefauthors}%
\unskip\
\newblock
\APACrefYearMonthDay{2025}{}{}.
\newblock
{\BBOQ}\APACrefatitle {A practical review of mechanistic interpretability for transformer-based language models} {A practical review of mechanistic interpretability for transformer-based language models}.{\BBCQ}
\newblock
\APACjournalVolNumPages{arXiv preprint arXiv:2407.02646}{}{}{}.
\PrintBackRefs{\CurrentBib}

\bibitem [\protect \citeauthoryear {%
Rohekar%
, Gurwicz%
\BCBL {}\ \BBA {} Nisimov%
}{%
Rohekar%
\ \protect \BOthers {.}}{%
{\protect \APACyear {2023}}%
}]{%
rohekar2023causal}
\APACinsertmetastar {%
rohekar2023causal}%
\begin{APACrefauthors}%
Rohekar, R\BPBI Y.%
, Gurwicz, Y.%
\BCBL {}\ \BBA {} Nisimov, S.%
\end{APACrefauthors}%
\unskip\
\newblock
\APACrefYearMonthDay{2023}{}{}.
\newblock
{\BBOQ}\APACrefatitle {Causal interpretation of self-attention in pre-trained transformers} {Causal interpretation of self-attention in pre-trained transformers}.{\BBCQ}
\newblock
\APACjournalVolNumPages{Advances in Neural Information Processing Systems}{36}{}{31450--31465}.
\PrintBackRefs{\CurrentBib}

\bibitem [\protect \citeauthoryear {%
Schick%
\ \protect \BOthers {.}}{%
Schick%
\ \protect \BOthers {.}}{%
{\protect \APACyear {2023}}%
}]{%
schick2023toolformer}
\APACinsertmetastar {%
schick2023toolformer}%
\begin{APACrefauthors}%
Schick, T.%
, Dwivedi-Yu, J.%
, Dess{\`\i}, R.%
, Raileanu, R.%
, Lomeli, M.%
, Hambro, E.%
\BDBL {}Scialom, T.%
\end{APACrefauthors}%
\unskip\
\newblock
\APACrefYearMonthDay{2023}{}{}.
\newblock
{\BBOQ}\APACrefatitle {Toolformer: Language models can teach themselves to use tools} {Toolformer: Language models can teach themselves to use tools}.{\BBCQ}
\newblock
\APACjournalVolNumPages{Advances in Neural Information Processing Systems}{36}{}{68539--68551}.
\PrintBackRefs{\CurrentBib}

\bibitem [\protect \citeauthoryear {%
Schneider%
}{%
Schneider%
}{%
{\protect \APACyear {2025}}%
}]{%
schneider2025generative}
\APACinsertmetastar {%
schneider2025generative}%
\begin{APACrefauthors}%
Schneider, J.%
\end{APACrefauthors}%
\unskip\
\newblock
\APACrefYearMonthDay{2025}{}{}.
\newblock
{\BBOQ}\APACrefatitle {Generative to Agentic AI: Survey, Conceptualization, and Challenges} {Generative to agentic ai: Survey, conceptualization, and challenges}.{\BBCQ}
\newblock
\APACjournalVolNumPages{arXiv preprint arXiv:2504.18875}{}{}{}.
\PrintBackRefs{\CurrentBib}

\bibitem [\protect \citeauthoryear {%
Shui%
\ \BBA {} Ru%
}{%
Shui%
\ \BBA {} Ru%
}{%
{\protect \APACyear {2025}}%
}]{%
shui2025bridging}
\APACinsertmetastar {%
shui2025bridging}%
\begin{APACrefauthors}%
Shui, X.%
\BCBT {}\ \BBA {} Ru, Z.%
\end{APACrefauthors}%
\unskip\
\newblock
\APACrefYearMonthDay{2025}{}{}.
\newblock
{\BBOQ}\APACrefatitle {Bridging the Gap Between Explainability and Large Language Models} {Bridging the gap between explainability and large language models}.{\BBCQ}
\newblock
\APACjournalVolNumPages{Hal-05011844}{}{}{}.
\PrintBackRefs{\CurrentBib}

\bibitem [\protect \citeauthoryear {%
Singhal%
\ \protect \BOthers {.}}{%
Singhal%
\ \protect \BOthers {.}}{%
{\protect \APACyear {2025}}%
}]{%
singhal2025toward}
\APACinsertmetastar {%
singhal2025toward}%
\begin{APACrefauthors}%
Singhal, K.%
, Tu, T.%
, Gottweis, J.%
, Sayres, R.%
, Wulczyn, E.%
, Amin, M.%
\BDBL {}others%
\end{APACrefauthors}%
\unskip\
\newblock
\APACrefYearMonthDay{2025}{}{}.
\newblock
{\BBOQ}\APACrefatitle {Toward expert-level medical question answering with large language models} {Toward expert-level medical question answering with large language models}.{\BBCQ}
\newblock
\APACjournalVolNumPages{Nature Medicine}{}{}{1--8}.
\PrintBackRefs{\CurrentBib}

\bibitem [\protect \citeauthoryear {%
Subramanian%
, Elango%
\BCBL {}\ \BBA {} Gungor%
}{%
Subramanian%
\ \protect \BOthers {.}}{%
{\protect \APACyear {2025}}%
}]{%
subramanian2025small}
\APACinsertmetastar {%
subramanian2025small}%
\begin{APACrefauthors}%
Subramanian, S.%
, Elango, V.%
\BCBL {}\ \BBA {} Gungor, M.%
\end{APACrefauthors}%
\unskip\
\newblock
\APACrefYearMonthDay{2025}{}{}.
\newblock
{\BBOQ}\APACrefatitle {Small Language Models (SLMs) Can Still Pack a Punch: A survey} {Small language models (slms) can still pack a punch: A survey}.{\BBCQ}
\newblock
\APACjournalVolNumPages{arXiv preprint arXiv:2501.05465}{}{}{}.
\PrintBackRefs{\CurrentBib}

\bibitem [\protect \citeauthoryear {%
Sun%
, Chen%
, Kolter%
\BCBL {}\ \BBA {} Liu%
}{%
Sun%
\ \protect \BOthers {.}}{%
{\protect \APACyear {2024}}%
}]{%
sun2024massive}
\APACinsertmetastar {%
sun2024massive}%
\begin{APACrefauthors}%
Sun, M.%
, Chen, X.%
, Kolter, J\BPBI Z.%
\BCBL {}\ \BBA {} Liu, Z.%
\end{APACrefauthors}%
\unskip\
\newblock
\APACrefYearMonthDay{2024}{}{}.
\newblock
{\BBOQ}\APACrefatitle {Massive activations in large language models} {Massive activations in large language models}.{\BBCQ}
\newblock
\APACjournalVolNumPages{arXiv preprint arXiv:2402.17762}{}{}{}.
\PrintBackRefs{\CurrentBib}

\bibitem [\protect \citeauthoryear {%
Taghanaki%
\ \BBA {} Monteiro%
}{%
Taghanaki%
\ \BBA {} Monteiro%
}{%
{\protect \APACyear {2025}}%
}]{%
taghanaki2025explain}
\APACinsertmetastar {%
taghanaki2025explain}%
\begin{APACrefauthors}%
Taghanaki, S\BPBI A.%
\BCBT {}\ \BBA {} Monteiro, J.%
\end{APACrefauthors}%
\unskip\
\newblock
\APACrefYearMonthDay{2025}{}{}.
\newblock
{\BBOQ}\APACrefatitle {Explain-Query-Test: Self-Evaluating LLMs Via Explanation and Comprehension Discrepancy} {Explain-query-test: Self-evaluating llms via explanation and comprehension discrepancy}.{\BBCQ}
\newblock
\APACjournalVolNumPages{arXiv preprint arXiv:2501.11721}{}{}{}.
\PrintBackRefs{\CurrentBib}

\bibitem [\protect \citeauthoryear {%
Tariq%
, Singh%
, Chhetri%
, Nepal%
\BCBL {}\ \BBA {} Paris%
}{%
Tariq%
\ \protect \BOthers {.}}{%
{\protect \APACyear {2025}}%
}]{%
tariq2025bridging}
\APACinsertmetastar {%
tariq2025bridging}%
\begin{APACrefauthors}%
Tariq, S.%
, Singh, R.%
, Chhetri, M\BPBI B.%
, Nepal, S.%
\BCBL {}\ \BBA {} Paris, C.%
\end{APACrefauthors}%
\unskip\
\newblock
\APACrefYearMonthDay{2025}{}{}.
\newblock
{\BBOQ}\APACrefatitle {Bridging Expertise Gaps: The Role of LLMs in Human-AI Collaboration for Cybersecurity} {Bridging expertise gaps: The role of llms in human-ai collaboration for cybersecurity}.{\BBCQ}
\newblock
\APACjournalVolNumPages{arXiv preprint arXiv:2505.03179}{}{}{}.
\PrintBackRefs{\CurrentBib}

\bibitem [\protect \citeauthoryear {%
{The Chairs and the Vice-Chairs of the General-Purpose AI Code of Practice}%
}{%
{The Chairs and the Vice-Chairs of the General-Purpose AI Code of Practice}%
}{%
{\protect \APACyear {2025}}%
}]{%
chairs2025third}
\APACinsertmetastar {%
chairs2025third}%
\begin{APACrefauthors}%
{The Chairs and the Vice-Chairs of the General-Purpose AI Code of Practice}.%
\end{APACrefauthors}%
\unskip\
\newblock
\APACrefYearMonthDay{2025}{}{}.
\newblock
\APACrefbtitle {Third Draft of the General-Purpose AI Code of Practice published, written by independent experts.} {Third draft of the general-purpose ai code of practice published, written by independent experts.}
\PrintBackRefs{\CurrentBib}

\bibitem [\protect \citeauthoryear {%
Triguero%
, Molina%
, Poyatos%
, Del~Ser%
\BCBL {}\ \BBA {} Herrera%
}{%
Triguero%
\ \protect \BOthers {.}}{%
{\protect \APACyear {2024}}%
}]{%
triguero2024general}
\APACinsertmetastar {%
triguero2024general}%
\begin{APACrefauthors}%
Triguero, I.%
, Molina, D.%
, Poyatos, J.%
, Del~Ser, J.%
\BCBL {}\ \BBA {} Herrera, F.%
\end{APACrefauthors}%
\unskip\
\newblock
\APACrefYearMonthDay{2024}{}{}.
\newblock
{\BBOQ}\APACrefatitle {General Purpose Artificial Intelligence Systems (GPAIS): Properties, definition, taxonomy, societal implications and responsible governance} {General purpose artificial intelligence systems (gpais): Properties, definition, taxonomy, societal implications and responsible governance}.{\BBCQ}
\newblock
\APACjournalVolNumPages{Information Fusion}{103}{}{102135}.
\PrintBackRefs{\CurrentBib}

\bibitem [\protect \citeauthoryear {%
Wang%
\ \protect \BOthers {.}}{%
Wang%
\ \protect \BOthers {.}}{%
{\protect \APACyear {2024}}%
}]{%
wang2024comprehensive}
\APACinsertmetastar {%
wang2024comprehensive}%
\begin{APACrefauthors}%
Wang, F.%
, Zhang, Z.%
, Zhang, X.%
, Wu, Z.%
, Mo, T.%
, Lu, Q.%
\BDBL {}others%
\end{APACrefauthors}%
\unskip\
\newblock
\APACrefYearMonthDay{2024}{}{}.
\newblock
{\BBOQ}\APACrefatitle {A comprehensive survey of small language models in the era of large language models: Techniques, enhancements, applications, collaboration with llms, and trustworthiness} {A comprehensive survey of small language models in the era of large language models: Techniques, enhancements, applications, collaboration with llms, and trustworthiness}.{\BBCQ}
\newblock
\APACjournalVolNumPages{arXiv preprint arXiv:2411.03350}{}{}{}.
\PrintBackRefs{\CurrentBib}

\bibitem [\protect \citeauthoryear {%
Wu%
\ \protect \BOthers {.}}{%
Wu%
\ \protect \BOthers {.}}{%
{\protect \APACyear {2024}}%
}]{%
wu2403usable}
\APACinsertmetastar {%
wu2403usable}%
\begin{APACrefauthors}%
Wu, X.%
, Zhao, H.%
, Zhu, Y.%
, Shi, Y.%
, Yang, F.%
, Liu, T.%
\BDBL {}others%
\end{APACrefauthors}%
\unskip\
\newblock
\APACrefYearMonthDay{2024}{}{}.
\newblock
{\BBOQ}\APACrefatitle {Usable XAI: 10 strategies towards exploiting explainability in the LLM era.} {Usable xai: 10 strategies towards exploiting explainability in the llm era.}{\BBCQ}
\newblock
\APACjournalVolNumPages{arXiv preprint arXiv:2403.08946}{}{}{}.
\PrintBackRefs{\CurrentBib}

\bibitem [\protect \citeauthoryear {%
C.~Xu%
, Guan%
, Greene%
, Kechadi%
\BCBL {}\ \protect \BOthers {.}}{%
C.~Xu%
\ \protect \BOthers {.}}{%
{\protect \APACyear {2024}}%
}]{%
xu2024contam}
\APACinsertmetastar {%
xu2024contam}%
\begin{APACrefauthors}%
Xu, C.%
, Guan, S.%
, Greene, D.%
, Kechadi, M.%
\BCBL {}\ \BOthersPeriod {.}\end{APACrefauthors}%
\unskip\
\newblock
\APACrefYearMonthDay{2024}{}{}.
\newblock
{\BBOQ}\APACrefatitle {Benchmark data contamination of large language models: A survey} {Benchmark data contamination of large language models: A survey}.{\BBCQ}
\newblock
\APACjournalVolNumPages{arXiv preprint arXiv:2406.04244}{}{}{}.
\PrintBackRefs{\CurrentBib}

\bibitem [\protect \citeauthoryear {%
Z.~Xu%
, Song%
\BCBL {}\ \BBA {} Lee%
}{%
Z.~Xu%
\ \protect \BOthers {.}}{%
{\protect \APACyear {2025}}%
}]{%
xu2025confronting}
\APACinsertmetastar {%
xu2025confronting}%
\begin{APACrefauthors}%
Xu, Z.%
, Song, T.%
\BCBL {}\ \BBA {} Lee, Y\BHBI C.%
\end{APACrefauthors}%
\unskip\
\newblock
\APACrefYearMonthDay{2025}{}{}.
\newblock
{\BBOQ}\APACrefatitle {Confronting verbalized uncertainty: Understanding how LLM’s verbalized uncertainty influences users in AI-assisted decision-making} {Confronting verbalized uncertainty: Understanding how llm’s verbalized uncertainty influences users in ai-assisted decision-making}.{\BBCQ}
\newblock
\APACjournalVolNumPages{International Journal of Human-Computer Studies}{}{}{103455}.
\PrintBackRefs{\CurrentBib}

\bibitem [\protect \citeauthoryear {%
Zhao%
\ \protect \BOthers {.}}{%
Zhao%
\ \protect \BOthers {.}}{%
{\protect \APACyear {2024}}%
}]{%
zhao2024explainability}
\APACinsertmetastar {%
zhao2024explainability}%
\begin{APACrefauthors}%
Zhao, H.%
, Chen, H.%
, Yang, F.%
, Liu, N.%
, Deng, H.%
, Cai, H.%
\BDBL {}Du, M.%
\end{APACrefauthors}%
\unskip\
\newblock
\APACrefYearMonthDay{2024}{}{}.
\newblock
{\BBOQ}\APACrefatitle {Explainability for large language models: A survey} {Explainability for large language models: A survey}.{\BBCQ}
\newblock
\APACjournalVolNumPages{ACM Transactions on Intelligent Systems and Technology}{15}{2}{1--38}.
\PrintBackRefs{\CurrentBib}

\bibitem [\protect \citeauthoryear {%
Zhu%
\ \protect \BOthers {.}}{%
Zhu%
\ \protect \BOthers {.}}{%
{\protect \APACyear {2024}}%
}]{%
zhu2024explanation}
\APACinsertmetastar {%
zhu2024explanation}%
\begin{APACrefauthors}%
Zhu, Z.%
, Chen, H.%
, Ye, X.%
, Lyu, Q.%
, Tan, C.%
, Marasovi{\'c}, A.%
\BCBL {}\ \BBA {} Wiegreffe, S.%
\end{APACrefauthors}%
\unskip\
\newblock
\APACrefYearMonthDay{2024}{}{}.
\newblock
{\BBOQ}\APACrefatitle {Explanation in the era of large language models} {Explanation in the era of large language models}.{\BBCQ}
\newblock
\BIn{} \APACrefbtitle {Proceedings of the 2024 Conference of the North American Chapter of the Association for Computational Linguistics: Human Language Technologies (Volume 5: Tutorial Abstracts)} {Proceedings of the 2024 conference of the north american chapter of the association for computational linguistics: Human language technologies (volume 5: Tutorial abstracts)}\ (\BPGS\ 19--25).
\PrintBackRefs{\CurrentBib}

\end{thebibliography}

\end{document}